\documentclass[a4paper,9pt,twocolumn]{extarticle} 
\usepackage[super,sort&compress,comma]{natbib} 
\usepackage[version=3]{mhchem} 
\usepackage[left=1.5cm, right=1.5cm, top=1.785cm, bottom=2.0cm]{geometry}
\usepackage{balance} 
\usepackage{graphicx} 
\usepackage{lastpage}
\usepackage[format=plain,justification=justified,singlelinecheck=false,font={stretch=1.125,small,sf},labelfont=bf,labelsep=space]{caption}
\usepackage{float} 
\usepackage[english]{babel}
\addto{\captionsenglish}{%
  
}
\usepackage{array} 
\usepackage{charter} 
\usepackage[T1]{fontenc} 
\usepackage{xcolor} 
\usepackage{setspace} 
\usepackage[compact]{titlesec}
\usepackage{hyperref} 
\usepackage{amsmath,amssymb} 
\usepackage[normalem]{ulem} 

\definecolor{cream}{RGB}{222,217,201}

\newcommand{\barri}[1]{\textcolor{black}{#1}}
\newcommand{\Barri}[1]{\textcolor{black}{#1}}

\begin{document}


\makeatletter
\renewcommand\LARGE{\@setfontsize\LARGE{15pt}{17}}
\renewcommand\Large{\@setfontsize\Large{12pt}{14}}
\renewcommand\large{\@setfontsize\large{10pt}{12}}
\renewcommand\footnotesize{\@setfontsize\footnotesize{7pt}{10}}
\makeatother

\renewcommand{\thefootnote}{\fnsymbol{footnote}}
\renewcommand\footnoterule{\vspace*{1pt}%
\color{cream}\hrule width 3.5in height 0.4pt \color{black}\vspace*{5pt}} 
\setcounter{secnumdepth}{5}

\makeatletter 
\renewcommand\@biblabel[1]{#1}            
\renewcommand\@makefntext[1]%
{\noindent\makebox[0pt][r]{\@thefnmark\,}#1}
\makeatother 
\renewcommand{\figurename}{\small{Fig.}~}
\setstretch{1.125} 
\setlength{\skip\footins}{0.8cm}
\setlength{\footnotesep}{0.25cm}
\setlength{\jot}{10pt}
\titlespacing*{\section}{0pt}{4pt}{4pt}
\titlespacing*{\subsection}{0pt}{15pt}{1pt}








\setlength{\arrayrulewidth}{1pt}
\setlength{\columnsep}{6.5mm}
\setlength\bibsep{1pt}

\makeatletter 
\newlength{\figrulesep} 
\setlength{\figrulesep}{0.5\textfloatsep} 

\newcommand{\topfigrule}{\vspace*{-1pt}%
\noindent{\color{cream}\rule[-\figrulesep]{\columnwidth}{1.5pt}} }

\newcommand{\botfigrule}{\vspace*{-2pt}%
\noindent{\color{cream}\rule[\figrulesep]{\columnwidth}{1.5pt}} }

\newcommand{\dblfigrule}{\vspace*{-1pt}%
\noindent{\color{cream}\rule[-\figrulesep]{\textwidth}{1.5pt}} }

\makeatother

\twocolumn[
\begin{@twocolumnfalse}




\vspace{1em}
\sffamily


\noindent\LARGE{\textbf{
Sedimentation and structure of squirmer suspensions under gravity 
}} \\


 \noindent\large{C. Miguel Barriuso G.$^{\ast}$\textit{$^{a,b\ddag}$} Horacio Serna,\textit{$^{a,b\ddag}$} 
 Ignacio Pagonabarraga,\textit{$^{c}$} 
 and Chantal Valeriani\textit{$^{a,b}$}} \\

\noindent\normalsize{The effect of gravity on the collective motion of living microswimmers, such as bacteria and micro-algae, is pivotal to unravel not only bio-convection patterns but also  the settling  of bacterial biofilms on solid surfaces.  
In this work, we investigate suspensions of microswimmers under the influence of a gravitational field and hydrodynamics, simulated via dissipative particle dynamics (DPD) coarse-grained model. 
We first study the collective sedimentation of passive colloids and microswimmers of the puller and pusher types upon increasing the imposed gravitational field and compare with previous results. Once  sedimentation occurs,
we observe that, as the gravitational field increases, the bottom layer undergoes a transition to an ordered state compatible with a hexagonal crystal. In comparison with passive colloids, both pullers and pushers easily rearrange at the bottom layer to anneal   defects. 
Specifically, pullers are better than pushers in preserving the hexagonal order of the bottom mono-layer at high gravitational fields.} \\

\end{@twocolumnfalse} \vspace{0.6cm}

]


\renewcommand*\rmdefault{bch}\normalfont\upshape
\rmfamily
\section*{}
\vspace{-1cm}



\footnotetext{\textit{$^{\ddag}$ The authors equally contributed to this work.}}

\footnotetext{\textit{$^{a}$~Departamento de Estructura de la Materia, Física Térmica y Electrónica, Universidad Complutense de Madrid, 28040 Madrid, Spain}}

\footnotetext{\textit{$^{b}$ GISC - Grupo Interdisciplinar de Sistemas Complejos 28040 Madrid, Spain}}

\footnotetext{\textit{$^{c}$ Departament de F\'isica de la Matèria  Condesada, Facultat de F\'isica - Universitat de Barcelona, Carrer de Mart\'i i Franquès, 1, 11, 08028 Barcelona, Spain  and Universitat de Barcelona Institute of Complex Systems (UBICS), Universitat de Barcelona, 08028 Barcelona, Spain}}




\section{Introduction}

In the last decade, many research efforts have been put forward trying to understand the effects of external fields of different kinds on the motion of active agents, paying special attention to cases in which the hydrodynamic interactions between active agents are relevant\cite{gompper20202020,marchetti2013hydrodynamics,stark2016swimming}. When acting concomitantly, external fields and hydrodynamics might induce new motion patterns in microswimmers. In particular, the influence of gravity on the collective motion of biological microswimmers, such as bacteria and micro-algae, is determinant to understand bio-convection patterns\cite{javadi2020photo,bees2020advances, pedley1992bioconvection,alloui2007numerical, ramamonjy2023pattern,kage2013drastic}, as well as the settling and growth of bacterial biofilms on solid surfaces \cite{lynch2006escherichia,mclean2001bacterial, korber1990effect}. When dealing with synthetic systems, it has been shown that confinement plus gravity plays a crucial role in the motion of catalytic Janus microswimmers, for which understanding its influence is key to controlling their trajectories\cite{ebbens2018catalytic,sharan2022upstream,singh2018photogravitactic,bailey2022microswimmers,bailey2024minimal}, \barri{or in experiments dealing with active droplets\cite{Kruger_maass_2016,thutupalli_flow-induced_2018,hokmabad_spontaneously_2022}, where clusters present a rich dynamics.} The collective motion of microswimmers affected by gravity  has promising technological applications such as stirring of bioreactors \cite{carvajal2024towards}, transport of fluid and small objects at the micro-scale \cite{dervaux2017light}, biomedicine \cite{chen2021overview}, and water remediation \cite{fu2022microscopic}.

A topic that has recently attracted attention is the sedimentation of active agents, being of interest not only the single-agent  but also the collective dynamics. 

The sedimentation dynamics of a single squirmer at high P\'eclet numbers under the influence of gravity is considered in ref. \citenum{ruhle2018gravity}, where  Multi Particle Collision Dynamics (MPCD)  was used. In this work the authors found different regimes such as cruising and sliding states, and proposed an expression for the sedimentation velocity as a function of  height,  self-propulsion and rotational velocities. Using the same model as in the present work, a recent numerical and experimental study 
revealed that the complex dynamics of a single Janus catalytic microswimmer swimming near a wall can be captured by a minimal squirmer-like model,  including bottom heaviness and mass asymmetry \cite{bailey2024minimal}.  By means of Dissipative Particle Dynamics (DPD), it was shown that \barri{the dynamics resulting  from the interplay between hydrodynamics and chemical fields 
in this type of system is overcame by inertial and gravitational effects}.  
A recent numerical work has focused on  the effects of the microswimmer's shape and vertical walls on the sedimentation dynamics 
\cite{ying2024study}: a two dimensional elliptical squirmer under a gravitational field is simulated inside a vertical channel employing the Lattice Boltzmann (LB) method. The observed dynamics, consisting of tilted motion and oscillations, depends on both the swimming mode (pusher, neutral or puller) and the squirmers' aspect ratio. The sedimentation dynamics of two squirmers of similar  and different types has also been studied in narrow vertical channels using the LB method \cite{nie2023two,guan2022swimming}, finding several motion patterns like steady and oscillatory settling.

Regarding collective sedimentation dynamics, hydrodynamic interactions have been considered in simulations of squirmer suspensions under the LB scheme \cite{alarcon2013spontaneous}, revealing that active stresses are responsible for flocking, swimming coherence and formation of band-like structures similar to those observed in systems of self-propelled particles with explicit alignment interactions \cite{vicsek1995novel,chate2008modeling,chate2020dry,kursten2020dry,solon2015phase}. 
LB simulations of squirmer suspensions with an additional tumbling rate and under the action of gravitational fields \cite{scagliarini2022hydrodynamic} have demonstrated  that for strong enough pullers/pushers, the hydrodynamic flows produced by their collective motion affect the sedimentation in a way 
that cannot be explained by the behavior of a single swimmer under gravity. 
On the other side, using MPCD, the authors of ref. \citenum{kuhr2017collective} have observed that most of the squirmers settle under a moderate gravitational field, forming a multi-layered structure at the bottom. Whereas squirmers on the surface of the sedimented layers  form convection patterns.  The authors found that the sedimentation length depends on the gravitational field applied and the squirmer type. 
In similar simulation studies of  bottom heavy squirmers, complex convective patterns were found \cite{ruhle2020emergent}, due to the  torque orienting  the squirmers upwards. The same behaviour was observed in gyrotactic clusters \cite{ruhle2022gyrotactic}. The influence of biological microswimmers (\textit{E. Coli}) on the settling of passive colloids has revealed that  the bacterial active bath enhances the effective diffusion coefficient of the colloidal particles \cite{maldonado2024sedimentation}.
 
Following the MPCD method, the collective dynamics of a monolayer of squirmers under strong gravitational fields was also investigated, finding a complex dynamical phase behaviour in terms of the swimming mode and the packing fraction of the monolayer that includes a hydrodynamic Wigner fluid, fluctuating chains, and swarms \cite{kuhr2019collective}. MPCD has also been used to study the sedimentation of attractively interacting colloids\cite{moncho2010effects}, determining that  clustering induced by strong attractions modified the linear relation between the sedimentation velocity and colloids' packing fraction. In the same spirit, experiments with colloidal beads\cite{lattuada2016colloidal} not only onserved the same  non-linear behaviour  but also reported that attractive interactions increase particle settling.

Apart from these studies on sedimented monolayers, few works have been devoted to suspensions of microswimmers under tight confinement in slit pores. The restriction to quasi two-dimensional motion promoted structural features similar to those observed in monolayers under strong gravity.
Reference \citenum{delfau2016collective} 
studied the effects of confinement in thin slit pores on the collective behavior of microswimmer suspensions, finding that the most relevant variable was the self-propelling mechanism, and suggesting that to improve the prediction of collective states the near-field approximation of the flow field should be considered. In a recent numerical article, the authors investigated  the combined effects of tight confinement and (LB) hydrodynamics  on the collective motion of microswimmers, concluding that the patterns largely differed  from their bulk counterparts \cite{bardfalvy2024collective}. 

In recent years considerable effort has been devoted to clarify how hydrodynamics affects the interaction of a microswimmer with bounding walls, taking into account the influence of swimmer pairwise/collective interactions and the walls boundary conditions. In particular, some studies show that pullers aggregate more at walls when solving Navier-Stokes equations with finite volume methods\cite{li_hydrodynamic_wall_2014}, considering stability arguments using boundary element methods\cite{ishimoto_squirmer_boundary_2013}, analyzing squirmer detention times at walls with MPCD simulations\cite{schaar_detention_2015} and more recently considering pair-wise interactions\cite{thery_lauga_maas_2023}. On the other side, works based on LB methods\cite{llopis_hydrodynamic_wall_2010,shen2018hydrodynamic} and a recent study (DPD-based)  of ellipsoidal microswimmers confined in thin films\cite{wu2024collective} have shown an opposite trend with pushers being effectively more attracted to walls than pullers. Depending on the volume fraction and the squirmer type (puller, neutral or pusher), swimmers exist in  a gas-like phase, undergo  swarming or Motility Induced Phase Separation. 


In the present work, we consider the collective sedimentation of squirmer-like microswimmers embedded in DPD solvent following an improved version of the model presented 
in ref. \citenum{barriuso2022simulating}. Apart from gaining insights on the interplay between gravity, thermal fluctuations and hydrodynamics, we aim to validate the model  contrasting it with previous simulation studies using different techniques \cite{kuhr2017collective, scagliarini2022hydrodynamic}. Once the sedimentation has occurred, we study the structure of the formed bottom layer, paying special attention to the transition to an ordered structure with hexagonal symmetry as the gravitational field increases. Although previous studies have also shown hexagonal order in a monolayer of squirmers \cite{kuhr2019collective}, we show that this order is compatible with a hexagonal crystal and that activity favours the repairing of defects observed in sedimented mono-layers of passive colloids.



We organize the article as follows. In section 2, we introduce the model, the simulation details and the analysis tools. In section 3, we first present the results of the sedimentation of microswimmers, and then we characterize the structure of the formed bottom monolayer of microswimmers. In section 4, we summarize the main findings of the article and present the conclusions.

\section{Numerical details}

\begin{figure*}[h!]
 \centering
 \includegraphics[width=\textwidth]{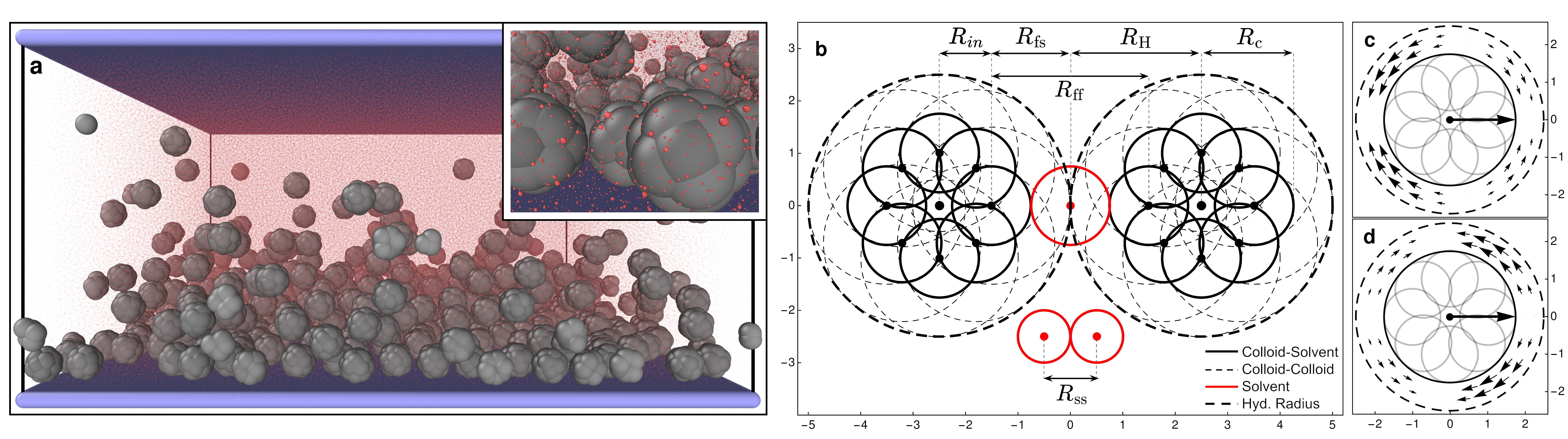}
 \caption{{\bf{(a)}} Snapshot of the full simulation box (inset: zoom on some bottom colloids). The raspberry colloids (gray) over the bounding substrate made of frozen particles (blue) surrounded the solvent particles (transparent red). See the supplementary material for an animated version of this figure (\texttt{video1.avi}). {\bf{(b)}} Schematic representation of a 2D section of two raspberry colloids (black) and three solvent particles (red). We see 8 of the 18 DPD \textit{filler} particles ---distributed on the surface of a sphere of radius $R_{in}=1.0R_{\text{ss}}$--- each colloid is made of. The filler particles have two DPD cutoffs, $R_{\text{ff}}=3$ when interacting with other filler particles (solid thick  circles), and $R_{\text{fs}}=R_{\text{ff}}/2$ when interacting with solvent particles (dashed thin circles). This allows solvent particles to flow between colloids while these are interacting and results in effective radius of $R_\text{cs}\equiv R_\text{c} = 1.75$ for colloid-solvent interactions and $R_\text{cc}\equiv R_\text{H} = 2.5$ for colloid-colloid interactions. Solvent particles interact with each other with a DPD cutoff $R_{\text{ss}}=1$. The interactions with the walls are also modelled as DPD with $R_{\text{wf}}=R_{\text{ff}}$ for wall-filler interactions and $R_{\text{ws}}=R_{\text{fs}}$ for wall-solvent. A \textit{thruster} particle (not visible), placed at the center of the colloid, ``generates'' the force fields shown in panels {\bf{(c)}} and {\bf{(d)}} in the region between spheres of radii $R_\text{c}$ and $R_\text{H}$. These emanate from the center of mass of the colloid, are fixed with its internal frame of reference, and are consistent with a \textit{pusher} (c) and a \textit{puller} (d) type squirmer, producing a net self-propulsion force along the axis of symmetry of the force field. Graphics presented here and elsewhere were generated in part using the visualisation software Ovito \cite{stukowski2009visualization}.}
 \label{fgr:Colloid}
\end{figure*}

\subsection{Model and simulation details}
\label{sec:sim_details}
The system under study is composed of $N=500$ ``raspberry'' colloids (swimming or not), composed of $N_c = 19$ particles each, embedded in a solvent composed of $N_s = 948073$ particles and confined between two parallel walls containing $N_w=168200$ particles, in the presence of a gravitational field only affecting the colloids (see fig. \ref{fgr:Colloid} a). The system is simulated with an in-house extension\footnote{\url{https://gitlab.com/hyperactivematter/acdpdh}} of the open source package LAMMPS \cite{plimpton1995fast} (\texttt{patch\_2Jul2021}) implementing colloid self-propulsion via a squirmer-like force field applied to the solvent particles, as in ref.\citenum{barriuso2022simulating}. All particles present in the system interact via DPD pairwise interactions (\texttt{DPD-BASIC} package\footnote{\url{https://docs.lammps.org/Packages_details.html\#pkg-dpd-basic}}) except particles composing the colloids which are joined together by rigid interactions (\texttt{RIGID} package\footnote{\url{https://docs.lammps.org/Packages_details.html\#pkg-rigid}}). The minimum length scale that we take as unit reference for all lengths present in the system is the DPD cutoff for solvent-solvent interactions $R_{\text{ss}}=1$ (see fig. \ref{fgr:Colloid}). The dimensions of the simulation box are $L_x = L_y = 87 R_{\text{ss}}$ and $L_z = 43.5 R_{\text{ss}}$. Periodic boundary conditions are applied along $x$ and $y$ axes. The solvent number density  $\rho_s=N/(V_{\text{box}}-V_{\text{cols}})=N/[L_x L_y L_z - N(4/3)\pi R_{\text{c}}^3]=2.981$, and the colloid volume fraction is $\phi_c = N\,(4/3)\pi R_{H}^3\,/\, L_x L_y L_z \approx 0.099$, where $R_{\text{c}}=1.75R_{\text{ss}}$ and $R_H=2.5 R_{\text{ss}}$ are the colloid radii with respect to the DPD cutoff for colloid-solvent and colloid-colloid interactions  respectively (see fig. \ref{fgr:Colloid} b).

\subsubsection{DPD forces}
The DPD force acting on all particles can be expresed as,
\begin{equation}
    \textbf{F}^\text{DPD}(\textbf{r}_{ij})= \left( F_{ij}^C + F_{ij}^D + F_{ij}^R \right)  \hat{\textbf{r}}_{ij} \quad \quad \text{if} \quad r_{ij}<r_c,
\end{equation}
being $\hat{\textbf{r}}_{ij}=(\textbf{r}_i-\textbf{r}_j)/r_{ij}$  the unit  vector between the $i$-th and $j$-th particles. The conservative term is $F_{ij}^C = A \, w(r_{ij})$, where $A$ is the amplitude and $w(r_{ij})=1 -r_{ij}/r_c$ a weighting factor varying between 0 and 1\cite{dpd_warren-groot}. The dissipative contribution reads $F_{ij}^D = - \gamma \, w^2(r_{ij}) \, \left(\hat{\textbf{r}}_{ij} \cdot \vec{\textbf{v}}_{ij} \right)$ with friction coefficient $\gamma$. Finally, the thermal contribution  $F_{ij}^R = \sigma \, \alpha \, w(r_{ij}) \, /\sqrt{\Delta t}$ is a random force, where  $\alpha$ is a Gaussian random number with zero mean and unit variance, $\Delta t$ the chosen time-step for the time integration and $\sigma = \sqrt{2 \,k_B T\, \gamma }$ is related to the mean of the random force via fluctuation-dissipation, being $T$ the temperature of the system. The values of the chosen DPD parameters are show in table \ref{tab:dpd_param}.

\begin{table}[h!]
\centering
\begin{tabular}{| c | c c c c c c|} 
 \hline
  & \texttt{ss} & \texttt{ff} &  \texttt{fs} & \texttt{ws} & \texttt{wf} & \texttt{t*}\\ [0.5ex] 
 \hline
 $A$ & $20$ & $50$ & $20$ & $150$ & $150$ & 0\\  
 $\gamma$ & $200$ & $0$ & $200$ & $450$ & $0$ & 0\\ 
 $R$ & $1$ & $3$ & $1.5$ & $1.5$ & $3$ & 0\\
 $k_BT$ & $1$ & $1$ & $1$ & $1$ & $1$ & 0\\\hline
\end{tabular}
\caption{Chosen values for the DPD parameters for the interactions between solvent (\texttt{s}), filler (\texttt{f}), wall (\texttt{w}) and thruster (\texttt{t}) particles. We have renamed $r_c\equiv R$ for notational convenience.}
\label{tab:dpd_param}
\end{table}

The chosen values  ensure low Reynolds number ($\text{Re}\le 0.105$), the lowest slip boundary conditions, impermeability of the walls and colloids by solvent particles and avoidance of depletion interactions between colloids and between colloids and walls. \footnote{Pathological depletion interactions stemming from the coarse-grained nature of DPD are well known and have been studied and tackled in different ways from the origins of the DPD framework until very rencently\cite{dpd_depletion_boek1998,dpd_depletion_gibson1999,dpd_depletion_whittle2001,dpd_depletion_whittle2010,dpd_depletion_barcelos2021,dpd_depletion_curk2024}.} 
\textcolor{black}{To minimize the impact of these interactions, we prescribe the different DPD cutoffs between interactions of solvent, colloidal and wall particles  to allow solvent to flow in between colloids or between colloids and walls; see fig.~\ref{fgr:Colloid}.}
To the best of our knowledge, this approach has not been yet studied, although it bears similarities with previous DPD studies, using virtual walls\cite{dpd_depletion_gibson1999} or dissipative ``coats''\cite{dpd_depletion_whittle2010}, and MPCD approaches using virtual particles for modelling flow past obstacles\cite{lamura_mpcd_2001} and squirmers\cite{zottl_simulating_2018}.

\subsubsection{Colloids}
Similarly to an early colloidal DPD model\cite{dpd_depletion_boek1998} the microswimmer body is modeled as a rigid-body raspberry\cite{Lobaskin_2004_raspberry} structure composed of 19 particles: 1 central particle (thruster) and 18 particles evenly distributed on a spherical surface of radius $R_{in} = 1.0R_{\text{ss}}$ (fillers). Further details of the model can be found in ref.~\citenum{barriuso2022simulating}. \Barri{In what follows we will indistinguishably refer  to  ``colloids'', ``microswimmers'' or ``squirmers'' as, although some differences exist,  the presented model can be applied to self-propelling bodies whether synthetic or biological\cite{shen2018hydrodynamic,campbell_experimental_2019}.}

Regarding the self-propulsion, we implement a variation of the squirmer model \cite{lighthill1952squirming,blake1971spherical} where, instead of prescribing a velocity boundary condition at the microswimmer surface, we apply a force field acting on the solvent particles located within a spherical shell surrounding the microswimmer. The resulting hydrodynamic force field only considering the two first surface modes of the polar component  (neglecting the radial component) reads,
\begin{equation}
    \textbf{F}^{H}(r,\theta) = \left( B_1 \sin{\theta} + B_2\sin{\theta}\cos{\theta}  \right)\hat{\textbf{e}}_{\theta} \quad \text{if}\quad R_c<r<R_H 
    \label{eq:SquirmerForce}
\end{equation}
where $r$ is the distance from the thruster to the solvent particle and  $\theta$ is the angle between the orientation of the microswimmer, $\hat{\textbf{e}}$,  and the solvent position vector. The tangential unit vector with respect to the colloid frame of reference is $\hat{\textbf{e}}_{\theta}$, and $R_c$ and $R_H$ define a shell around the colloid where the force is applied (see fig. \ref{fgr:Colloid}). The self-propulsion of a given microswimmer originates when the solvent particles, over which the hydrodynamic force field is distributed, exert an opposite reaction force to the nearest filler particle of the microswimmer (see ref.~\citenum{barriuso2022simulating} for more details). 

The parameter $B_1$ determines the propulsion force along the prescribed orientation of the colloid, and $B_2$ accounts for the swimming mode. The sign of the parameter $\beta = B_2/B_1$ (determined by that of $B_2$) defines the self-propulsion mechanism: pusher ($\beta<0$), neutral ($\beta = 0$) or puller ($\beta > 0$). 
\textcolor{black}{Rather than prescribing the propulsion force, $F_p$, we prescribe $B_1$ and compute the propulsion force $F_p = 2B_1/3$ through the functional relation for the squirmer velocity.} A suspension of passive colloids, $B_1 = 0$, will be used as a reference system.

The colloid is not a perfect sphere and steric effects might modulate the orientational behavior of the colloid at close interactions (see SM). Larger sphericity can be achieved by increasing the number of fillers. A perfect spherical colloid with  steric interactions is achievable with the current model considering  the conservative interaction 
of only the central (thruster) particle, whose DPD cutoff would be increased as to include  the remaining 18 filler particles. The   conservative interactions of the fillers  have  to be switched off, only  needed for interchanging dissipative, thermal and thrust forces with the rest of the solvent particles. Even though  simulations of more complex colloid geometries might be not as direct as with the raspberry model,   both the model and the DPD scheme are versatile enough to consider either spherical swimmers or swimmers with complex structures  in ref. \cite{barriuso2022simulating}.

\subsubsection{Solvent  hydrodynamics}
\label{s:SolventHydro}

\barri{The maximum Reynolds number is estimated as $\text{Re} = v_p \,R_{\text{c}}\, /  \, \nu \approx 0.105$, where $R_c=1.75$ is the effective solvent colloid radius  (see fig. \ref{fgr:Colloid}), $\nu = 45k_B T \, / \, 4\pi\gamma_{\text{ss}}\rho_s R_{\text{ss}}^3\,+\,2\pi\gamma_{\text{ss}}\rho_s R_{\text{ss}}^5/1575$ is the estimated kinematic viscosity~\cite{dpd_warren-groot} and $v_p$ is the colloid Stokes' propulsion velocity $v_p=F_p\,/\,6\pi\eta R_c$, being  $\eta$ the dynamic viscosity $\eta = \rho_s\nu$. The largest P\'eclet number   is set to $\text{Pe}=v_p \, R_c \, / \, D_{\text{col}}^0 = 58.33$ corresponding to a propulsion force $F_p= 100/3$. The passive colloid diffusion coefficient is estimated as $D_{\text{col}}^0 = k_B T \,/\, 6\pi\eta R_c$.}

\barri{The implementation of the DPD boundary conditions  is a relevant and widely discussed topic 
~\cite{revenga_boundary_1998,willemsen_no-slip_2000,pivkin_new_2005,mehboudi_systematic_2011,ranjith_no-slip_2013,li_dpd_complex_geometries,wang_new_2021}.  
Following 
ref.~\citenum{willemsen_no-slip_2000},  we increase  the wall density  (4 times the solvent density) and independently tune 
the solvent-wall particles friction coefficient  $\gamma_{\text{ws}}$  
(to ensure the lowest slip velocity and no density fluctuations near the wall, we separately launch simulations  of steady solvent flow). 
However, 
in the simulations presented here (where a steady flow is not present) 
a 
slip velocity was measured for passive colloids and walls (see SM).}

\barri{As in any other coarse-grained method, we capture  hydrodynamics  down to the coarse-graining scale $R_{\texttt{ss}}$.
\textcolor{black}{This approach describes near field hydrodynamics~\cite{ishikawa_hydrodynamic_2006,gotze_mesoscale_2010,spagnolie_hydrodynamics_2012,lintuvuori_hydrodynamic_2016}, but  does not capture the divergence  of lubrication at contact. The role of the asymptotic  lubrication and far-field interactions 
has been previously addressed  for pairs of squirmers~\cite{ishikawa_development_2008, delfau2016collective, ruhle2018gravity}. Specifically, when}
interacting with walls, the far-field approximation predicts velocities that agree within a 10\% error up to a distance of $0.5$ body lengths from the wall\cite{spagnolie_hydrodynamics_2012}. In our case, using the data from the radial distribution functions (see SM), we estimate the minimum gap between the colloid and the wall to be  $\delta_w^{\text{RDF}} \approx 0.7 D_{\texttt{cs}}^{\text{RDF}}$ and between colloids $\delta_c^{\text{RDF}} \approx D_{\texttt{cs}}^{\text{RDF}}$, where $D_{\texttt{cs}}^{\text{RDF}}\approx 2 R_{\texttt{ss}}$ is the estimated minimum swimmer body length with respect to  solvent interactions\footnote{Since we choose a larger DPD filler-wall interactions cutoff  than the  filler-solvent one, there will always be a gap between the colloid-wall effective surfaces  with respect to the solvent interactions. 
If there was no overlap, this gap would be $\delta_w = R_{\texttt{wf}}-R_{\texttt{fs}}/2-R_{\texttt{ws}}/2=1.5R_{\texttt{ss}}=0.43D_{\texttt{cs}}$, where $D_{\texttt{cs}} = 2R_{\texttt{cs}}  = 2R_{in}+R_{\texttt{fs}}=3.5R_{ss}$ is the swimmers body-length with respect to  filler-solvent interactions (see fig. \ref{fgr:Colloid}). However, due to the softness of the DPD interaction, there is some overlap: thus, the minimum effective gap (estimated via the radial distribution functions, see SM) is  $\delta_w^{\text{RDF}} \approx 1.4R_{\texttt{ss}} = 0.7 D_{\texttt{cs}}^{\text{RDF}}$. Likewise, when colloids interact with each other, without overlap, the gap is $\delta_c = R_{\texttt{ff}}-R_{\texttt{fs}}=1.5R_{\texttt{ss}}=0.43D_{\texttt{cs}}$ and estimating distances via RDFs, $\delta_c^{\text{RDF}} \approx D_{\texttt{cs}}^{\text{RDF}}$. The reason the gap, measured in bodylengths, is larger when the RDFs are considered (which might seem counter-intuitive) is because the overlap between filler and solvent particles is greater than between fillers or between wall and filler particles, since $A_{\texttt{fs}}<A_{\texttt{ff}}<A_{\texttt{wf}}$ (see table \ref{tab:dpd_param}). Thus, the overlap in the measuring unit (the bodylength), $D_{\texttt{cs}}-D_{\texttt{cs}}^{\text{RDF}}=1.5R_{\texttt{ss}}$, is greater than in the measured quantities (the gaps), $\delta_c - \delta_c^{\text{RDF}}=0.5R_{\texttt{ss}}$ and $\delta_w - \delta_w^{\text{RDF}}=0.1R_{\texttt{ss}}$.}.}
\barri{The lubrication torque correction\cite{lintuvuori_hydrodynamic_2016,ruhle2018gravity} at the mentioned distance from the wall is found to be $\mathcal{T}\approx 50$, while the average torque felt by the colloids is $\approx 600$. 
\textcolor{black}{Therefore, although not  dominant, lubrication forces should be included to make accurate predictions.}
The aim of this work is not to make quantitative predictions  but to validate and asses how the model captures the known phenomenology, exploring its limitations when applied to current non trivial systems (sedimentation of a suspension of swimmers) and gaining insight into the behavior of such systems.}

\subsubsection{Walls}
Walls are placed at $z = l_w/2$ and $z = L_z - l_w/2$ and consist of frozen DPD beads describing a square lattice with lattice constant $l_w = 0.3 R_{\text{ss}}$, to achieve impermeability (no solvent particles crossing) and smoothness of the wall. 
The difference between the minimum and maximum DPD conservative forces along the wall surface at $z=1.35R_{\text{ss}}$ is $\Delta F<0.038$ (see SM). The DPD parameters for wall particles are reported in table \ref{tab:dpd_param}.

\subsubsection{Gravitational field}

A gravitational field of magnitude $F_g$,  only affecting the squirmers, is imposed along the $z$ axis. The gravitational force acting on each  colloid's 
particle (filler or thruster) is  given by,
\begin{equation}
    \textbf{F}^g = -mg\hat{\textbf{z}} 
\end{equation}
where $m=1$ is the mass, $g$ is the gravitational acceleration and $\hat{\textbf{z}}$ is the unit vector along the $z$ axis. As we  only apply this force to  colloids, buoyant forces are not present in the system. Thus, the sedimentation velocity does not depend on the density difference between the colloid and the solvent.   

\subsubsection{Equations of motion}
\barri{In summary, particles obey Newton's equations of motion}
\textcolor{black}{where the   total force acting on  particle $i$ of type $\alpha$ (as in table \ref{tab:dpd_param}) reads}
\begin{align}
    \textbf{F}_{i\alpha} &= 
    \sum_{j\beta \neq i\alpha}\textbf{F}^{\text{DPD}}_{i\alpha,j\beta}
    \left(1-\delta_{\beta\texttt{t}}\right)
    +
    \textbf{F}^{\text{H}}_{i\alpha,j\beta}\left(\delta_{\alpha\texttt{s}}\delta_{\beta\texttt{t}}-\delta_{\alpha\texttt{f}}\delta_{\beta\texttt{s}}\delta_{\langle i,j\rangle}\right)+\nonumber\\
    &+
    \textbf{F}^g_{i\alpha}\delta_{\alpha\texttt{f}},
\end{align}
\noindent where $\delta_{\alpha\beta}$ is Kronecker's delta and $\delta_{\langle i,j\rangle}$ is 1 if the $i$-th filler particle is the nearest to the $j$-th solvent particle and 0 otherwise (considering that multiple solvent particles can have the same nearest filler particle).

\subsubsection{Initialization}
Colloids are initialized in a $10\times 10\times 5$ cubic lattice 
with random orientations. Solvent is initialized randomly in the empty space left by the colloids and the walls. \textcolor{black}{Fixing the colloidal positions, switching  gravity and self-propulsion off,    the solvent   equilibrates 
for $5\cdot 10^3$ steps with $dt=0.01$. Subsequently, we   perform  production runs for $t_{\text{sim}}=5\cdot 10^5$ steps with  $dt=0.01$.}

\subsubsection{Units}
All quantities can be  expressed in reduced dimensionless units with 
the solvent-solvent DPD cutoff  as unit of length, $r_0=R_{\text{ss}}$, the mass of a DPD solvent bead as the mass unit $m_0=m_{\texttt{DPD}}$ and the thermal energy $\epsilon_0 = k_BT$. The time unit is  defined as $t_0=r_0/\sqrt{2\epsilon_0/m_0}$. 

\subsection{Analysis tools}
\subsubsection{Sedimentation}

\barri{To validate the colloidal raspberry-DPD  model under gravity, we first study  sedimentation of passive colloids.  Selecting only the 100 topmost initialized colloids, we track their positions from the initial configuration until they reach a stationary sedimented configuration. We compute their sedimentation velocity $v_g$ by a linear fit of their averaged trajectories for $z > 6 R_H$ and compare it with the Stokes law, $v_g = F_g\,/\,6\pi \eta R_c$, where $\eta$ is the   DPD dynamic viscosity  in section \ref{s:SolventHydro}. Fitting the $\{F_g,\,v_g\}$ data, we obtain an effective sedimentation radius, that we  compare with the radius estimated with the radial distribution function.}

\barri{Once  colloids have sedimented and the mean  colloids' height reached a stationary regime, we  compute their sedimentation profile via their 2D packing fraction in the $xy$-plane, together with  their orientational profile (i.e. their mean vertical orientation) along the $z$-axis. For the latter calculation, we divide the $z$ direction in 200 bins of height $\Delta z \approx 0.22R_{\text{ss}}$ and compute the number of colloids in each bin, together with  their mean vertical orientation. Finally, we take the temporal average over the last 400 configurations for $\tau > 3000$.  }
\begin{align}
\label{eq:density_2d_paching_fraction}
    \phi_{\text{\,2D}}(\Delta z_i) &= \frac{\pi R_H^2}{L_x L_y}\left\langle N(\Delta z_i)\right\rangle\\
\label{eq:orientational_profile}    
    \left\langle \cos\alpha \right\rangle(\Delta z_i) &= 
    \left\langle \frac{1}{N(\Delta z_i)} \sum_{j\in\Delta z_i}\hat{n}\cdot\hat{\textbf{e}}_j \right\rangle
\end{align}
\barri{Where $N(\Delta z_i)$ is the number of colloids whose centers of mass are contained in the $\Delta z_i$ bin, $\hat{n}$ is the normal vector from the bottom to the top wall, parallel to the $z$ axis unit vector and $\hat{\textbf{e}}$  the colloids orientation. }

\subsubsection{Analysis tools: Characterization of the bottom layer}
Once  colloids have sedimented, we study the structure of the bottom layer (\textcolor{black}{colloids for which $z < 
 2R_H$}) as a function of the imposed gravitational force, $F_g$. Averages are taken over $600$ independent configurations in steady state.

All reported  quantities refer to  colloids at the bottom layer.  
To characterize the structure of the bottom layer, we compute the pair correlation function, $g(r)$ and the static structure factor (SSF) in two dimensions

\begin{equation}
    S(\textbf{q}) = \left\langle \frac{1}{N}\sum_{i=1}^{N}\sum_{j=1}^{N}e^{-i\textbf{q}\cdot(\textbf{r}_i-\textbf{r}_j)}\right\rangle,
    \label{eq:SSF}
\end{equation}
here $\textbf{q} = (2\pi/L_x)(n_x,n_y)$ ($n_x$ and $n_y$ integers) is the wave-vector and $\textbf{r}_i$ is the center of mass  position of colloid $i$ projected onto the  $xy$ plane, and $N$ is the number of colloids at the bottom layer. To gain further information on the bottom layer's structure, we calculate the colloid  
hexagonal order parameter \cite{halperin1978theory,nelson1979dislocation}
\begin{equation}
    \psi_6(\textbf{r}_k) =\left \lvert \frac{1}{n_{b}} \sum_{j=1}^{n_b} e^{i6\theta_{kj}} \right \rvert, 
    \label{eq:psi_6}
\end{equation}
where $\theta_{kj}$ is the angle formed by the displacement vector between the centers of mass of colloids $k$ and $j$, $\textbf{r}_{kj} = \textbf{r}_j - \textbf{r}_k$ and the $x$ axis. The sum runs over the number of nearest neighbors, $n_b$, of colloid $k$ within a cutoff radius, $r_{\psi_{6}} = 6.0R_{ss}$, which correspond approximately to the first minimum of the radial distribution function at strong gravitational fields.
If a colloid $k$ has less than 2 closest neighbours, we assign $\psi_6(k) = 0$. Next, we compute the probability distribution of the hexagonal order parameter, $P(\psi_6)$ and the hexagonal correlation function,
\begin{equation}
    G_6(\textbf{r}) = \langle\psi_6(\textbf{r}_j)\psi_6^*(\textbf{r}_i)\rangle,
    \label{eq:G_6}
\end{equation}
to analyze the decaying of the hexagonal ordering. 
 
To study  colloids' alignment with respect to their closest neighbours, we compute each colloid $k$ local polar order parameter 
\begin{equation}
    \mathcal{P}_l(k) = \frac{1}{n_b} \left\lvert \sum _{i=1}^{n_b} \hat{\textbf{e}}_i\right\rvert,
    \label{eq:P_l}
\end{equation}
 where $\hat{\textbf{e}}_i$ is the prescribed orientation of the colloids along which the hydrodynamic force field is applied. The sum runs over the $n_b$ nearest neighbours within a cut-off radius, $r_{\mathcal{P}_l} = r_{\psi_6}$ for each colloid $k$, including itself. For isolated swimmers, $\mathcal{P_l} = 0$ Although $\mathcal{P}_l(k)$ is calculated 
 at the bottom layer, we consider the three-dimensional orientation of the colloids since they might point either towards the wall or away from the wall. 
Next, we compute the probability distribution of the local polar order parameter, $P(\mathcal{P}_l)$. The total alignment of the bottom layer can be described by the global polar order parameter\cite{vicsek1995novel}, 
\begin{equation}
    \mathcal{P}(t) = \left\langle\frac{1}{N} \left\lvert \sum _{i=1}^{N} \hat{\textbf{e}}_i\right\rvert \right \rangle
    \label{eq:GlobalP}
\end{equation}
To complete the analysis of colloids' alignment, we study the  spatial velocity correlation function (SVCF) and the spatial orientation correlation function (SOCF),
\begin{align}
    SVCF(\textbf{r}) = \left \langle \hat{\textbf{v}}_i \cdot \hat{\textbf{v}}_j \right \rangle 
    \label{eq:SVCF} \\ 
    SOCF(\textbf{r}) = \left \langle \hat{\textbf{e}}_i \cdot \hat{\textbf{e}}_j \right \rangle
    \label{eq:SOCF}
\end{align}
$\hat{\textbf{v}}_i$ and $\hat{\textbf{v}}_j$ are the three-dimensional unit velocities of particles $i$ and $j$, and $\hat{\textbf{e}}_i$ and $\hat{\textbf{e}}_j$ are the unit orientations. Thus, both correlations adopt values in the range $[-1, 1]$.

The presence of the wall influences the orientation of the colloids close to it. Hence, we define the angle $\alpha$ between the colloids orientation and the normal vector of the bottom wall, $\hat{n} = [0, 0, 1]$, and
calculate its distribution, $P(\alpha)$.

\section{Results and Discussion}
We first report briefly the results for sedimenting passive colloids and compare with previous knowledge. Once sedimentation takes place, we analyze the sedimentation profiles of puller/pusher squirmers and characterize the structural features of the sedimented bottom layer using the passive case as reference and comparing with previous studies.

\subsection{Sendimentation velocity of passive colloids}
\begin{figure}[h!]
    \centering
    \includegraphics[width=\columnwidth]{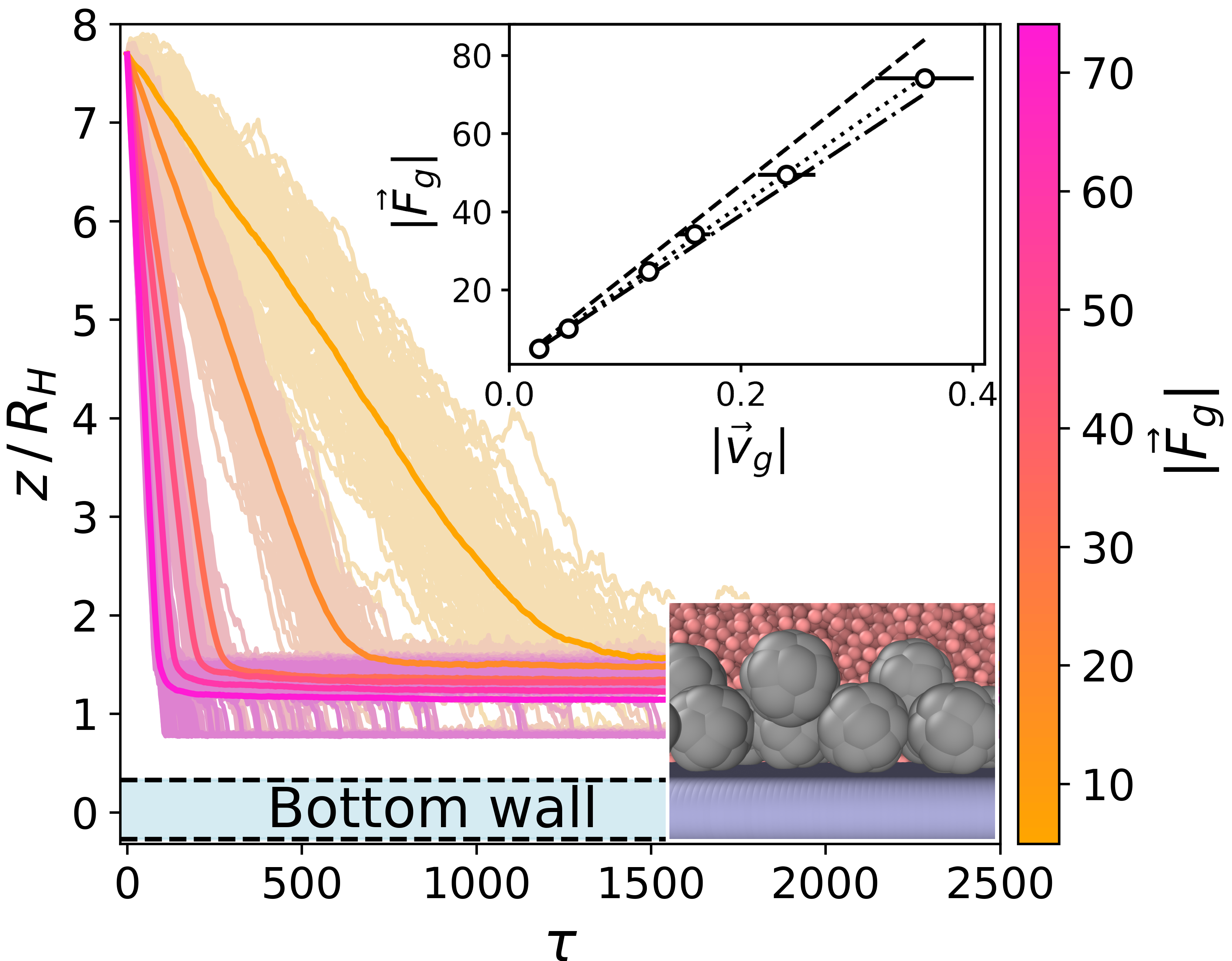}
    \caption{Trajectories for the sedimentation of the top layer of passive colloids (light colors) and average (dark colors) for  $F_g=\{5,10,25,34,50,74\}$. The $y$-axis is scaled to the body length of the colloids $D_H=2R_H$. The bounding wall is displayed in light blue at the bottom. Top inset: The data points correspond to the obtained sedimentation velocities by fitting the averaged $z$ trajectories (dark color curves), the error bars show the ensemble standard deviation. The dashed line corresponds to Stokes law for a colloid with radius $R=R_{in}+R_{\text{ss}}/2=1.75$ (see fig. \ref{fgr:Colloid}). The dotted line corresponds to a fit to the data points corresponding to  a colloid radius of $R=1.55$. The dashed-dotted line corresponds to Stokes law for a colloid with radius $R=1.46$ estimated from the RDF. Bottom inset: snapshot of the sedimented colloids at the highest gravity $F_g = 74.1$.}
    \label{fig:sedimentation}
\end{figure}

\barri{Passive colloids sediment falling,  on average, along linear trajectories in the vertical direction (see fig. \ref{fig:sedimentation}).} 
\barri{The measured sedimentation velocity, $v_g$, is  proportional to the  $F_g=6\pi\eta R_c\, v_g$} 
%
\barri{Increasing the packing fraction of the colloidal dispersion has the effect of decreasing the sedimentation velocity due to hydrodynamic interactions. Using  $R_{c,\text{RDF}}=1.46$ the packing fraction is  $\phi = N(4/3)\pi R_{c,\text{RDF}}^3/L_xL_yL_z\approx 0.02$. According to fig. 12.3 of ref. \citenum{Russel_1989_colloidal_dispersions}, such a packing fraction has the effect of  reducing the sedimentation velocity by $5\%$:  
in our case $100\cdot(R_{c,\text{eff}}-R_{c,\text{RDF}})/R_{c,\text{RDF}}\approx 6\%$.} It should be noted that hydrodynamic theory predicts a varying friction coefficient depending on the distance to the wall \Barri{(see refs. \citenum{perkins_hydrodynamic_1992,ruhle2018gravity} and eq. 7-2.15 of ref. \citenum{happel_low_1983}). The approximation  in ref. \citenum{ruhle2018gravity} for the system under study predicts a decrease in the sedimentation velocity of a single colloid,  although barely distorting the linearity of the trajectory for $10 \lesssim z/R_{\texttt{ss}} \lesssim 35$. 
The disentanglement of these two effects (the varying sedimentation velocity due to walls and due to collective sedimentation) lies outside the scope of the present work and is left for future study.} 

\subsection{Squirmer sedimentation and orientational profiles}
\label{sec:sedimentaion}

\begin{figure}[h!]
    \centering
    \includegraphics[width=\columnwidth]{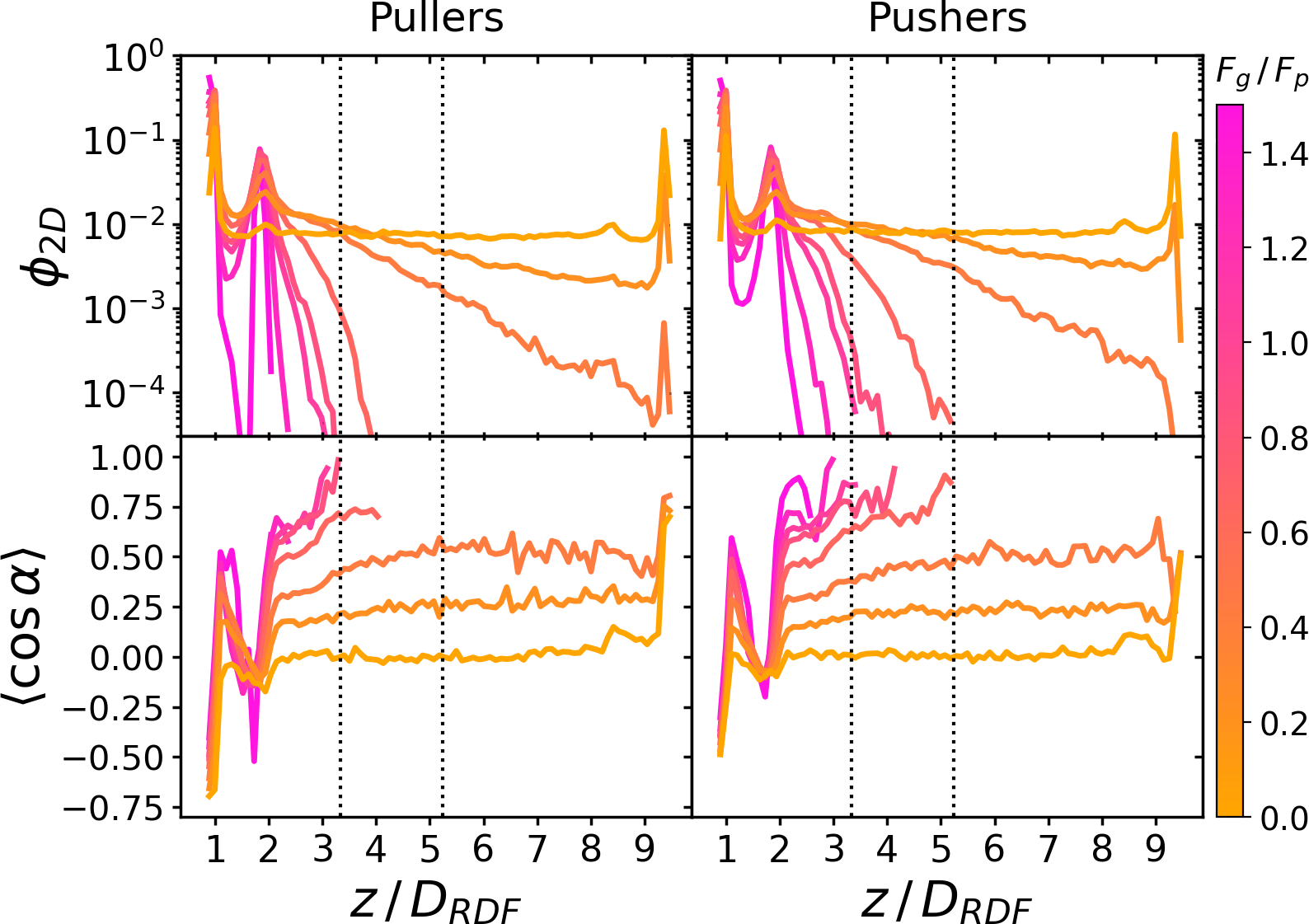}
    \caption{{\bf Top:} Sedimentation profile  as the 2D packing fraction $\phi_{\,\text{2D}}$ for pullers (left) and pushers (right) for $F_g/F_p = \{ 0.0,\,0.15,\,0.3,\,0.5,\,0.7,\,0.75,\,1.0,\,1.5 \}$ (color bar). The $x$-axis is scaled to the colloids body length $D_{RDF} = 2R_{RDF}\approx 4.2$ regarding colloid-colloid interactions estimated by the RDF (see SM). The region between the dotted lines corresponds to the exponential regime in  the two lowest nonzero $F_g/F_p$ ratios.
    {\bf Bottom:} Mean vertical orientation along the $z$ direction. $\alpha$ is the angle between the vertical and the squirmer orientation ($\cos\alpha =  1$ corresponds to an upward (positive $z$) orientation and $\cos\alpha =  -1$  to an downward (negative $z$) orientation). To be compared with fig. 3 of ref. \citenum{kuhr2017collective}. Check the SM for a zoom on the layers.}
    \label{fig:SedimentationProfiles}
\end{figure}

\Barri{Figure \ref{fig:SedimentationProfiles} reports the sedimentation and orientational profile of the squirmer dispersions.}

\textcolor{black}{In the absence of gravity, $F_g=0$, squirmers are uniformly distributed in the bulk, as expected.} Near the walls, the characteristic accumulation of active particles~\cite{rothschild_non-random_1963,berke_lauga_wall_attraction_2008,li_accumulation_2009,molaei_failed_2014,li_hydrodynamic_wall_2014} appears as two peaks at $z\approx D_{\text{RDF}}$ and $z\approx 9.4\,D_{\text{RDF}}$ reaching $\phi_{\text{2D}}\approx 0.14$ and $\phi_{\text{2D}}\approx 0.09$ for pullers and pushers,  respectively (see  SM for a zoom on the peaks). This  implies  that pullers tend to stick slightly more to walls than pushers. Since these packing fractions are not far from the dilute limit, if we assume that the difference in aggregation of pullers and pushers at the wall is mostly controlled  by  hydrodynamic interactions of a single squirmer with the wall, the difference in packing fractions qualitatively agrees  with the theory~\cite{lauga_hydrodynamics_2009,lauga_fluid_2020}: 
when swimming towards the bottom wall, $\alpha\in (\pi/2,3\pi/2)$, pullers tend to reorient towards the wall and align perpendicularly to it  ($\alpha = \pi$), while pushers tend to reorient away from the wall and align parallel to it ($\alpha = \pm\pi/2$).
\textcolor{black}{Thermal fluctuations on a pusher swimming parallel to the wall will favor with equal probability that the squirmer approaches or moves away from the wall. Since pullers are oriented preferentially towards the wall,  thermal fluctuations will not help the squirmer to displace away from the wall.}

\barri{As gravity increases, the packing fraction maximum of the first layer also increases, reaching $\phi_{\text{2D}}\approx 0.82$ for pullers and $\phi_{\text{2D}}\approx 0.78$ for pushers, being in all cases slightly higher for pullers than for pushers (see fig. 4 of SM)\footnote{It should be noted that these values were obtained computing the 2D packing fraction for the first layer with a  width bin size of $\Delta z = R_{RDF} = 2.1R_{ss}$. In SM-fig. 3  we study  the peaks for a smaller bin size ($\Delta z = 0.044 R_{ss}$): the peaks height's difference   between pullers and pushers is reduced as gravity increases.  The peaks' widths and tails  are also affected by  gravity; thus,  when we compute the packing fraction with $\Delta z = R_{RDF}$,  the difference between pullers and pushers appears constant.}. The same effect is more clearly visible at the top wall ($z\approx 9.4\,D_{\text{RDF}}$) for the $F_g/F_p=0.3$ case, where the pullers' packing fraction   presents a peak near the wall (that is absent for pushers).}

\barri{Away from the walls, at $z= 2\,D_{\text{RDF}}$ and $z=8.4\,D_{\text{RDF}}$, two secondary peaks are visible reaching $\phi_{\text{2D}}\approx 0.0050$ and $\phi_{\text{2D}}\approx 0.0057$ for pullers and pushers, respectively.  As gravity increases, the packing fraction of the second bottom layer for both pullers and pushers remains similar and increases linearly, until $F_g/F_p \approx 0.7$ when it reaches a maximum and starts decreasing (see SM). This maximum appears when the gravity is strong enough to confine all  colloids below $z=3\,D_{\text{RDF}}$. Thus, increasing gravity beyond this value does not increase the the second layer packing fraction, since there are no more colloids above $z=3\,D_{\text{RDF}}$. Instead, increasing gravity decreases the packing fraction, since colloids from the second layer are pushed down to the first, not yet fully populated. Additionally, since pushers can leave the first layer more easily than pullers\cite{wu2024collective}, as we increase gravity beyond this value, the packing fraction for pushers is higher than for pullers at all $0.7 < F_g/F_p < 1.5$ values. Thus, if gravity is large enough, it is slightly more probable to find pushers than pullers in the second layer, as opposed to what happens in the first sedimented layer.}

\barri{In the region between the first and the second layers we also observe a non-monotonic behaviour of the 2D packing fraction (see SM). It increases for pullers and pushers (being larger for pullers than for pushers) until it reaches a maximum around $F_g/F_p \approx 0.3$. Then, it starts decreasing for  pullers and pushers and starting from $F_g/F_p \ge 0.7$   a crossover appears (with  pushers overtaking pullers).}

Moving away from the bottom layers, $z > 2.5 D_{\text{RDF}}$, a decay of the 2D packing fraction is observed, always faster for pullers than for pushers. At the two lowest gravitational forces, this decay seems exponential:  
thus, the corresponding sedimentation lengths, $\delta$, is obtained by fitting the packing fraction to $\phi_{\,\text{2D}}=e^{\,z/\delta}$ between two chosen heights (the dotted lines in fig. \ref{fig:SedimentationProfiles}) 
\textcolor{black}{as described in ref.~\citenum{kuhr2017collective},  to minimize the influence of the  walls.}
The corresponding sedimentation lengths are (see SM),
\begin{align}
    &\delta_{\,\texttt{pull}}  = 5.27\,R_{\text{RDF}}, \quad &\delta_{\,\texttt{push}} = 9.96\,R_{\text{RDF}}\quad \text{for}\quad &F_g/F_p=0.15,  \nonumber\\
    &\delta_{\,\texttt{pull}} = 2.5\,R_{\text{RDF}},  \quad &\delta_{\,\texttt{push}} = 3.41\,R_{\text{RDF}}\quad \text{for}\quad &F_g/F_p=0.3, \nonumber
\end{align}
i.e. the sedimentation length decreases with increasing gravity and it is always higher for pushers than for pullers, agreeing with previous numerical\cite{kuhr2017collective,scagliarini2022hydrodynamic} and experimental\cite{palacci_sedimentation_2010,ginot_nonequilibrium_2015} results. Moreover,   no clear sign of bioconvection was found in this exponential regime. The reason might come from the geometry of the simulation box (wide and short height vs. the tall column used in ref. \citenum{kuhr2017collective}) and the fact that the signal of convection seems to become blurrier upon increasing the strength of the stresslet\cite{kuhr2017collective}, \Barri{consistently with a decreased collective behavior\cite{delfau2016collective}}
. Finally, 
for $F_g/F_p \ge 0.5$ some heights become unreachable, although pushers are able to access slightly higher regions than pullers.

\begin{figure*}[h!]
 \centering
 \includegraphics[width=12cm]{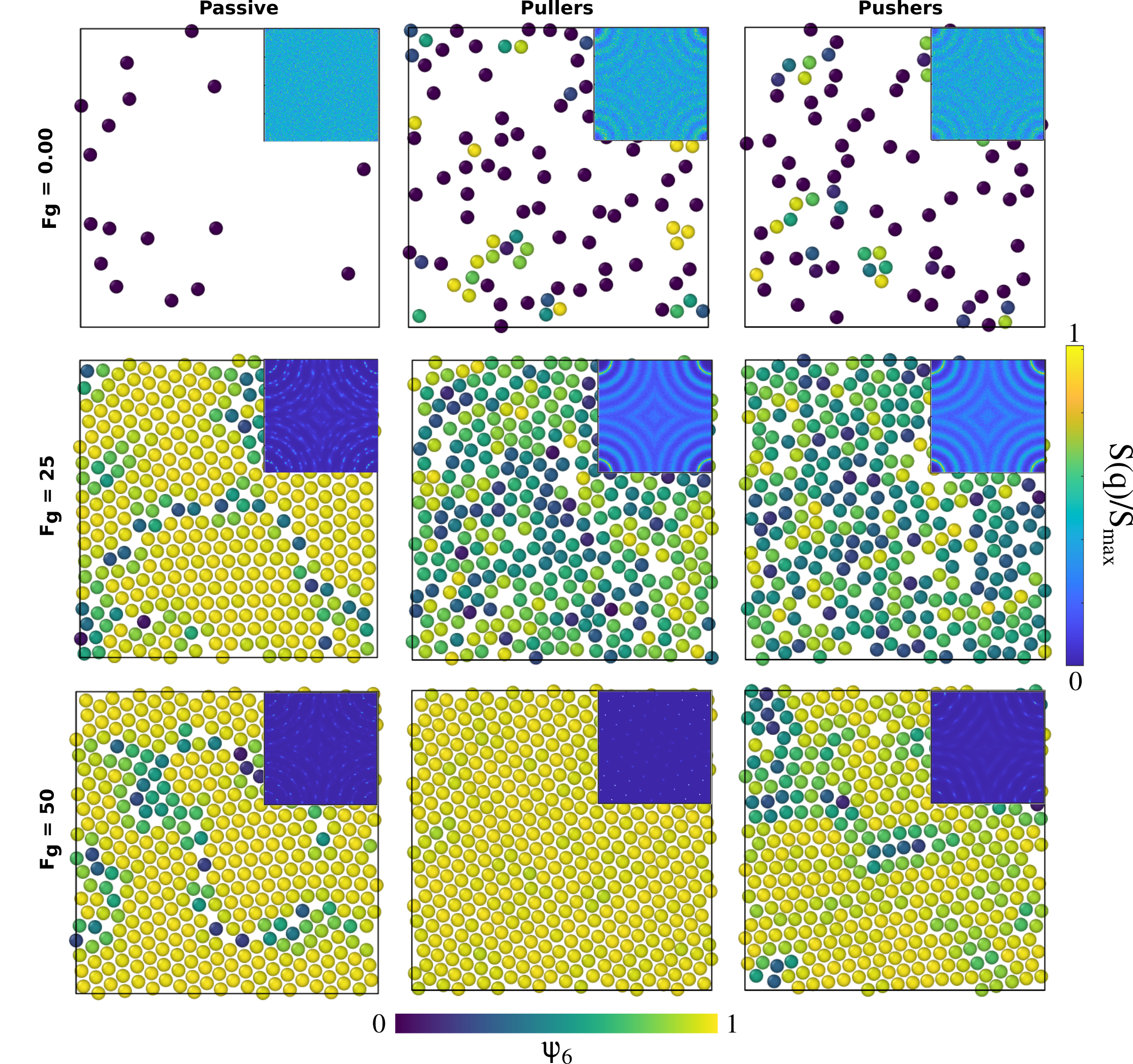}
 \caption{Configurations of the colloids' centres of mass  at the bottom layer in the steady state for passive, pullers and pushers at different  $F_g$. The microswimmers' propulsion force   is $F_p = 33.33$.  Colloids are represented with a radius of $R_{C, eff} = 1.85$, being $2R_{C,eff}$ the contact distance between colloids according to the pair correlation function  in the top row of fig. \ref{fgr:Correlations}. The colloids' color code   is the value of the hexagonal order parameter, $\psi_6$, as in the color bar. The insets are the two-dimensional Static Structure Factor (SFF) measured for the colloids at the bottom layer (the color bar  represents the normalized intensity, $S(q)/S_{max}$, of the SFF). The axes of the insets (not shown) are the components of the wave vector $q_x$ and $q_y$ and range from $-2\pi$ to $2\pi$. An animated version of this figure is provided in the SI.
 }
 \label{fgr:SSF2D}
\end{figure*}


%



\Barri{The bottom panels of fig. \ref{fig:SedimentationProfiles} show the  squirmers' mean vertical orientation  at different heights.} 
\textcolor{black}{ In the absence of gravity, $F_g=0$, the alignment is dominated by  steric and hydrodynamic interactions\cite{lauga_hydrodynamics_2009}, and thus   subject to the squirmers shape and type.} 
\textcolor{black}{Although the description of the squirmer through a raspberry model  does not correspond to a perfectly spherical shape, the deviations in sphericity it introduces, e.g. wall alignment due to the short range structure of the steric interaction,  are small (see SM). Moreover, when comparing the relative behavior of pullers and pushers, the  effects due to  weak departure from sphericity are the same in both,
so the differences in the observed behavior can be attributed solely to hydrodynamics.}
\textcolor{black}{In the far-field approximation}, a single squirmer interacting will tend to align perpendicularly or parallel  to a wall for a puller or pusher respectively. 
\Barri{The} $F_g/F_p=0$ curve of the bottom panels of fig. \ref{fig:SedimentationProfiles} 
\Barri{displays} two opposing peaks near the top and bottom walls at $z\approx D_{RDF}$ and $z\approx 9.4\,D_{RDF}$ reaching $\left\langle\cos\alpha\right\rangle\approx\pm 0.7$ for pullers and $\left\langle\cos\alpha\right\rangle\approx\pm 0.5$ for pushers ($\approx 135^{\circ}$ and $\approx 120^{\circ}$ w.r.t each wall normal respectively). 
\Barri{To gain further information, the distribution $P(\alpha)$ for the squirmers close to the bottom wall and the full joint probability distributions $P(z,\cos\alpha)$ also for higher $z$ are reported in the bottom panel of fig. \ref{fgr:Distributions} and in fig. 5 of the SM respectively.}
\textcolor{black}{They display }
the different shapes of the distributions for pullers and pushers and show that the most probable values near walls are $\cos\alpha\approx\pm 0.9$ and $\cos\alpha\approx\pm 0.4$ respectively ($\approx 155^{\circ}$ and $\approx 115^{\circ}$  w.r.t each wall normal). Therefore,  both pullers and pushers are oriented towards the wall but pullers are more tilted towards it. The detailed discussion of the results for squirmers close to the bottom wall ($z<1.5 D_{RDF}$) 
\Barri{will be addressed in} the next section (\S \ref{sec:bottom_layer}).
Two secondary peaks are observed around  $z= 2\,D_{\text{RDF}}$ and $z=8.4\,D_{\text{RDF}}$ (zoom avaible in SM) that also show a mean orientation slightly tilted towards the nearest wall $\left\langle\cos\alpha\right\rangle\approx\pm 0.1$.
\barri{Further into the bulk 
a constant $\left\langle\cos\alpha\right\rangle\approx 0$ \Barri{is observed} that stems from an isotropic distribution (see SM, first column of fig. 5).}

\barri{For the lowest gravities considered, $F_g/F_p \le 0.3$, the orientations in the bulk region tilt upwards for both pullers and pushers until the top wall is reached 
\textcolor{black}{as squirmers more aligned with the vertical direction can travel a wider span of  heights.} 
Although both pullers and pushers reach the top wall, only pullers display an upward peak at $z\approx 9.4\,D_{RDF}$ concurring with their tendency to reorient perpendicular to the wall via hydrodynamic interactions, \textcolor{black}{ in agreement with ref.~\citenum{kuhr2017collective}}. As  gravity increases, squirmers need a more upward orientation to access the same heights than for weaker gravity}. For $F_g/F_p = 1/2$, heights above $z\gtrsim 4 D_{RDF}$ and $z\gtrsim 5 D_{RDF}$ become unreachable for pullers and pushers respectively. 
\textcolor{black}{Pushers access slightly higher regions than pullers, since  they tend to align parallel to each other (decreasing $\alpha$) while pullers tend to} 
align perpendicularly (increasing  $\alpha$) thus sinking~\cite{thery_lauga_maas_2023}.
\textcolor{black}{Close to the second layer, at $z\approx 2D_{RDF}$,  a depletion as particles swim} away from the bottom wall is observed for all $F_g/F_p$. As we increase $z$, particles start reorienting upwards consistently with previous predictions\cite{enculescu_active_2011} (see also fig. 5 of SM for a more direct comparison).

\barri{Our results for $F_g/F_p=0.15$ are in agreement to those reported in    ref. \citenum{scagliarini2022hydrodynamic}, since  we also detect  a bimodal $P(z,\cos\theta)$  for close to the wall pullers, for which  negative orientations are more likely than positive ones (fig. 5 of SM, 2nd column). Whereas for pushers, $P(z,\cos\theta)$ is unimodal and more centered (in our case the most likely orientations do not exactly match  probably due to the colloid's asphericity). Our bulk distribution for pullers is also alike, with a noticeable bias towards positive orientations, although the one for pushers does not differ from pullers as much as those reported in ref.\citenum{scagliarini2022hydrodynamic}.}



\subsection{Characterization of the bottom layer}\label{sec:bottom_layer}

Figure \ref{fgr:SSF2D} represents the formation and structure of the bottom layer once  sedimentation has occurred 
\textcolor{black}{ for  $B_1 = 50$ and $\beta = \pm 10$, corresponding to  $F_p = 2B_1/3 = 100/3$, with $0.00\leq F_g\leq 75$} 
and the passive case as a reference system. 

Upon increasing gravity, $F_g$, in both active and passive systems,  colloids deposited at the bottom layer undergo a transition to a phase with hexagonal order, \textcolor{black}{which depends on the systems activity. The hydrodynamic  signature of the squirmers affects not only  when the transition takes place but also the structural properties of the crystalline  phase}. 
Steady state snapshots of the colloids at the bottom layer at three selected gravitational fields, $F_g$, are shown in fig. \ref{fgr:SSF2D}, where 
the colloids are represented in different colors depending on the values of their hexagonal order parameter, $\psi_6$. The insets of the snapshots contain the two-dimensional SSF (see Eq. \ref{eq:SSF}). 

In the absence of gravity, fig. \ref{fgr:SSF2D} top row,  the SSF of the passive system exhibits a rather homogeneous intensity, meaning that the structure is disordered and resembles a gas. The SSF for both pullers and pushers shows   signatures of a liquid structure but not fully developed. 
The almost-liquid structure of the active systems is also reflected in the fact that both pullers and pushers tend to accumulate close to the walls even without gravity, as reported in the packing fraction profiles of fig. \ref{fig:SedimentationProfiles}.
\textcolor{black}{Visual inspection of the colloids close to the wall (see supplementary videos) shows that as gravity increases}
the system undergoes a transition from a \textit{kissing}-like state\cite{kuhr2019collective} to a weak dynamic clustering phase. \textcolor{black}{Although  our model includes  near-field hydrodynamics,} 
\textcolor{black}{a similar behavior has been reported in models  only including  far-field and lubrication interactions~\cite{thery_lauga_maas_2023}.
At an intermediate gravity}, $F_g = 25$ (central row), passive colloids form a solid structure and the SFF shows features of a defective hexagonal crystal \textcolor{black}{as defects appear at short and long ranges.  Squirmers, both pullers and pushers,  display an SSF of a fully developed liquid.}
Finally, at a high gravitational field, $F_g = 50$ (bottom row),  all SFFs display the features of a hexagonal ordered phase, with  defects present for passive colloids and pushers, whereas pullers form an almost perfect hexagonal crystal. 
Once  passive colloids are deposited at the bottom, they do not change their positions 
due to thermal motion, and thus the bottom layer is kinetically trapped  as a crystal  with defects. For long enough simulations, the bottom layer of passive systems should reach a perfect hexagonal order. In contrast, due to  activity, the suspensions of pullers and pushers easily overcome the kinetically trapped states 
forming  an ordered hexagonal phase on a much shorter time-scale. From the snapshots and the SFF, it is already visible that pullers  tend to maintain the hexagonal order better than pushers. An animated version of fig. \ref{fgr:SSF2D} is provided in the SM.

\begin{figure}[h!]
 \centering
 \includegraphics[width=8.8cm]{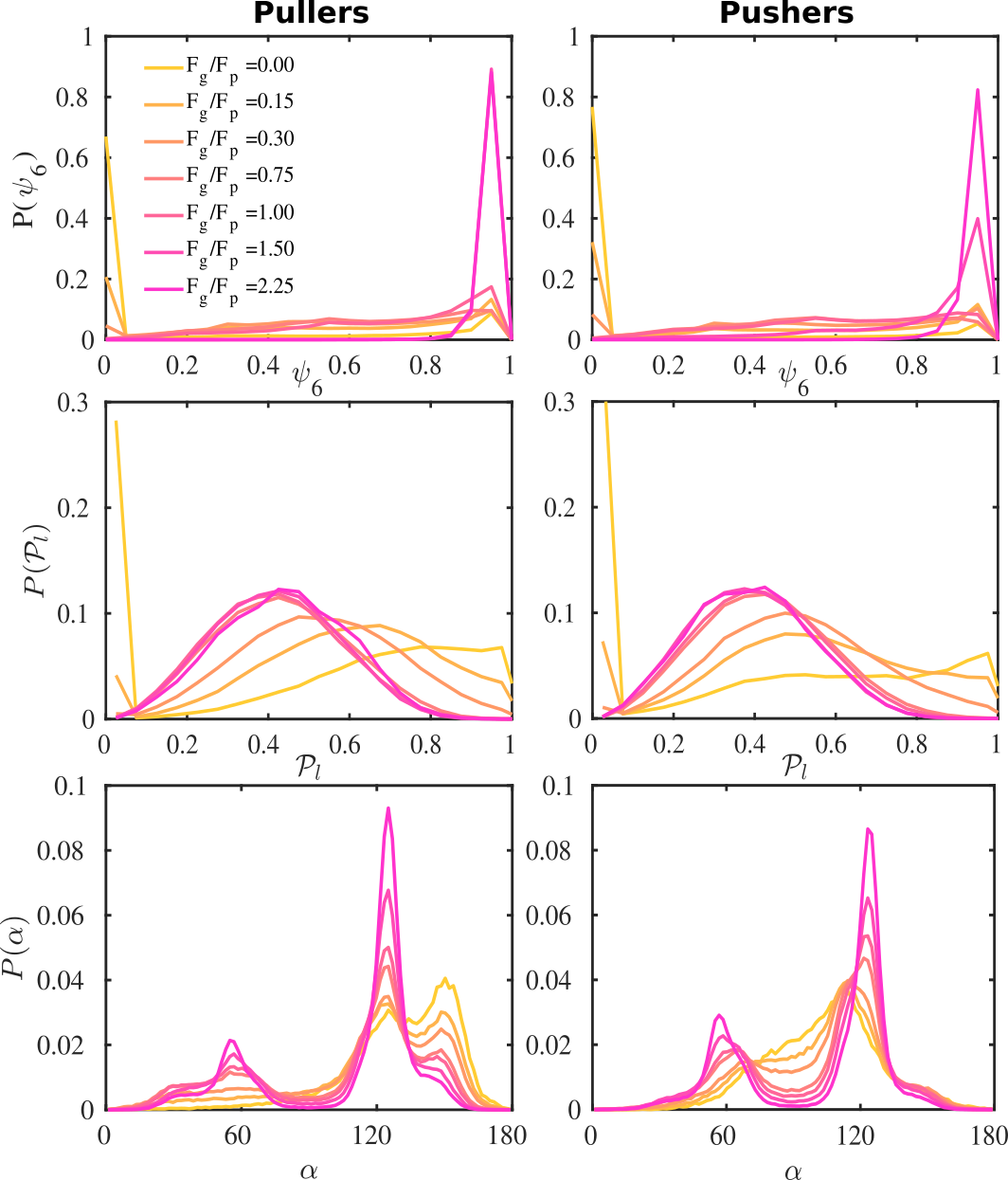}
 \caption{ Probability distribution functions of the hexagonal order parameter, $P(\psi_6))$ (top row), the local polar order parameter, $P(\mathcal{P}_l)$ (central row), and the angle between the prescribed orientation of the microswimmers and the bottom wall's normal vector, $P(\alpha)$ (bottom row), for puller ($\beta = 10$) and pushers ($\beta = -10$) at different  $F_g/F_p$. The bin size to calculate $P(\psi_6)$ and $P(\mathcal{P}_l)$ is $0.05$ and to calculate $P(\alpha)$ is $\pi/100$. See SM for $P(\alpha)$ distributions at different heights. 
 }
 \label{fgr:Distributions}
\end{figure}

To better understand  the formation of the hexagonal phase and its characteristics, we present in fig. \ref{fgr:Distributions} the probability distributions of the hexagonal order parameter (see Eq. \ref{eq:psi_6}), $P(\psi_6)$, the local polar order parameter (see Eq. \ref{eq:P_l}), $P(\mathcal{P}_l)$, and the angle between colloids' orientations and the normal vector of the bottom wall, $P(\alpha)$, for suspensions of pullers and pushers for all   $F_g/F_p$ values considered. 

\textcolor{black}{For weak gravities, $F_g/F_p <= 0.30$, }$P(\psi_6)$ (top row in fig.\ref{fgr:Distributions}) suspensions of  pullers and pushers  display a very similar behavior, featuring a peak around $\psi_6 = 0.0$ that decreases as $F_g$ increases. This peak is due to the presence of many isolated colloids with no neighbours for which hexagonal order cannot be defined and thus $\psi_6$ is assigned \textcolor{black}{ to zero. Otherwise,} the distributions are spread rather uniformly up to high values of $\psi_6$ where a small peak is already visible. For $F_g/F_p = 0.75$, there are enough colloids in the bottom layer and thus $\psi_6$ \textcolor{black}{is defined for all the squirmers.}
The distributions flatten and the small peak around high values of $\psi_6$ \textcolor{black}{decreases.}
At high gravities, ($F_g/F_p > 1.00$), the distributions for both pullers and pushers display a pronounced peak around $\psi_6 = 0.95$. Nevertheless, the distributions for pullers are virtually overlapping for the two highest gravities ($F_g/F_p = 1.50$, $2.25$), whereas the distributions for pushers indicate that the fully-developed hexagonal phase is only observed at the highest gravitational field of $F_g/F_p = 2.25$ and the distribution for $F_g/F_p = 1.50$ only shows a 
\textcolor{black}{small} peak. This points out that 
\textcolor{black}{pullers} favour the formation of hexagonal order and thus the transition \textcolor{black}{develops at smaller gravities than for pushers.} 

\begin{figure}[h!]
 \centering
 \includegraphics[width=5cm]{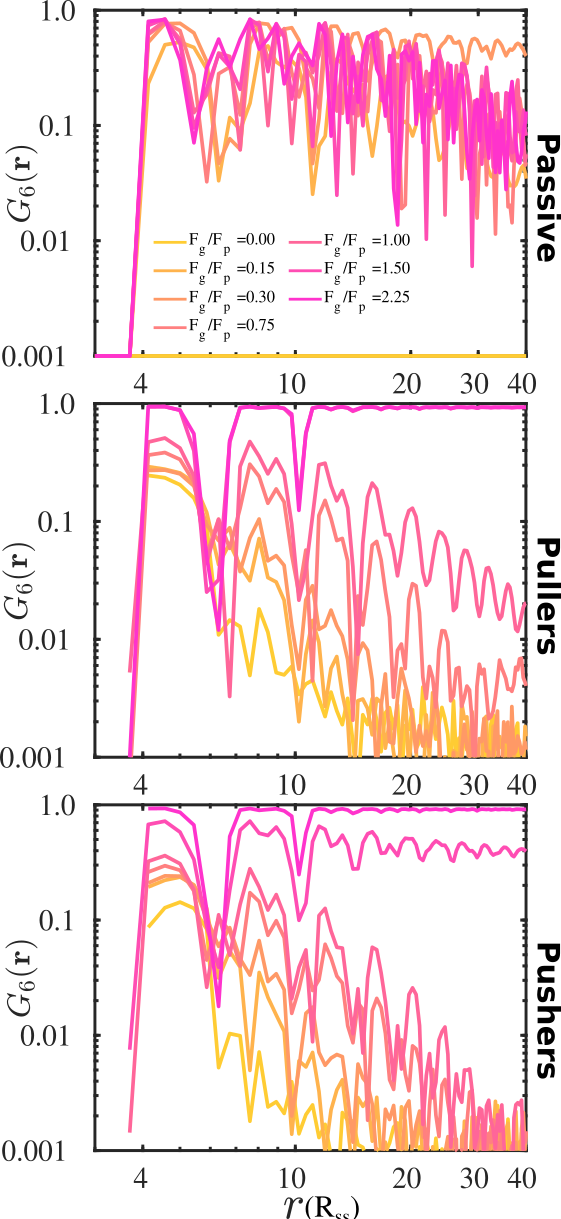}
 \caption{ The hexagonal order correlation functions, $G_6(\textbf{r})$, for passive colloids, pullers ($\beta = 10$), and pushers ($\beta = -10$), upon increasing the imposed gravitational field. The distance is in  units of the DPD cut-off solvent, $R_{ss}$, and the bin size used to calculate the correlations is $0.435R_{ss}$
 }
 \label{fgr:G6}
\end{figure}

To complement the analysis of the $P(\psi_6)$ and be able to qualitatively classify the hexagonal phases at the bottom layer, we compute the hexagonal order correlation function, $G_6(r)$ (see Eq.~\ref{eq:G_6}), and \textcolor{black}{display them} in fig. \ref{fgr:G6}. 
For passive colloids,  the correlations decay slowly and are noisy due to the defects on the hexagonal structure. For pullers, at high gravitational fields, 
\textcolor{black}{ $G_6(r)$ does} not decay, 
as expected  for a hexagonal crystal. At smaller gravitational fields,  \textcolor{black}{ $G_6(r)$ shows} 
the typical exponential decay of liquids. \textcolor{black}{ Pushers display the signature of a hexagonal crystal} 
for the highest
gravity, $F_g/F_p = 2.25$, and probably a power law decay indicating a hexatic phase for $F_g/F_p = 1.50$. At smaller gravitational fields, the correlations decay exponentially.

We now investigate the local polar order \barri{which has been observed in experiments of active droplets\cite{thutupalli_swarming_2011}, Janus particles\cite{nishiguchi_mesoscopic_2015} and even vibrated active disks\cite{deseigne_collective_2010} in which hydrodynamics is not present}. Computing the distributions of the local polar order parameter, $P(\mathcal{P}_l)$ (central row in fig.\ref{fgr:Distributions})  allows us to determine the alignment of colloids with respect to their first neighbors. \textcolor{black}{In the absence of gravity} 
$P(\mathcal{P}_l)$ 
displays  a pronounced peak around zero \textcolor{black}{ both for pullers and pushers, because} 
\textcolor{black}{ many colloids do not have neighbors.}
The distribution for pullers is  shifted to higher values of $\mathcal{P}_l$ than for pushers, \barri{ consistently with previous results\cite{oyama_purely_2016,evans_orientational_2011}}. Both distributions also feature a small peak around $\mathcal{P}_l = 0.95$, corresponding to the alignment of  colloids with few neighbors. 
\textcolor{black}{Local alignment decreases with gravity, leading to a left shift  of $\mathcal{P}_l $.} 
For $F_g/F_p = 0.15$ and $0.30$, $\mathcal{P}_l $ exhibits wide peaks 
at  $\mathcal{P}_l \approx 0.7$ and $0.5$, \textcolor{black}{for pullers and pushers respectively}, indicating, on average, an 
alignment between neighbors \textcolor{black}{coming from the local interactions between microswimmers. In the supplementary video 
\texttt{video4\_local-pol.avi} it can be observed how microswimmers (both pullers and pushers) within $r_{\mathcal{P}_l}$ align with themselves when gravity is absent. In the case of  pullers, the local alignment occurs when groups of swimmers form small clusters  rapidly disappearing due to  swimmers moving in different directions. In the case of  pushers, besides the formation of  high local polar-order clusters,  we often observe that, after an encounter,  swimmers  move together along the bottom wall during short times before moving apart.
To confirm that local polarization is not only produced by squirmers aligning with the wall but also by squirmers aligning with each other, 
we show the local polar order parameter distributions, $P\left(\mathcal{P}_{l,\text{2D}}\right)$, calculated using the 2D projections of the swimmers' orientation onto the bottom wall's plane (see fig. 6 of the SM). These display a peak around $0.95$, for both pullers and pushers, which is higher for pushers, consistently with their tendency to align parallel to the bottom wall and between each other. 
}
\textcolor{black}{For pushers, at $F_g/F_p = 0.15$ and $0.30$, the distributions feature peaks around $\mathcal{P}_l \approx 0.5$.
For $F_g/F_p >=0.75$, the distributions peak at $\mathcal{P}_l \approx 0.4$. As gravity becomes stronger, the 
packing fraction at the bottom layer increases and local alignment is hindered due to the increase of squirmer interactions that randomize the orientations because of the high strength of the stresslets. 
The resulting distribution approaches the expected one for uniformly distributed 3D orientations, 
although slightly shifted to higher $\mathcal{P}_l$ values because of the induced vertical alignment coming from the steric interactions due to the swimmers asphericity. This is further confirmed since the 2D distributions $P(\mathcal{P}_{l,2D})$ are in very good agreement with the expected one for uniformly distributed 2D orientations.}

\begin{figure}[h!]
 \centering
 \includegraphics[width=0.8\columnwidth]{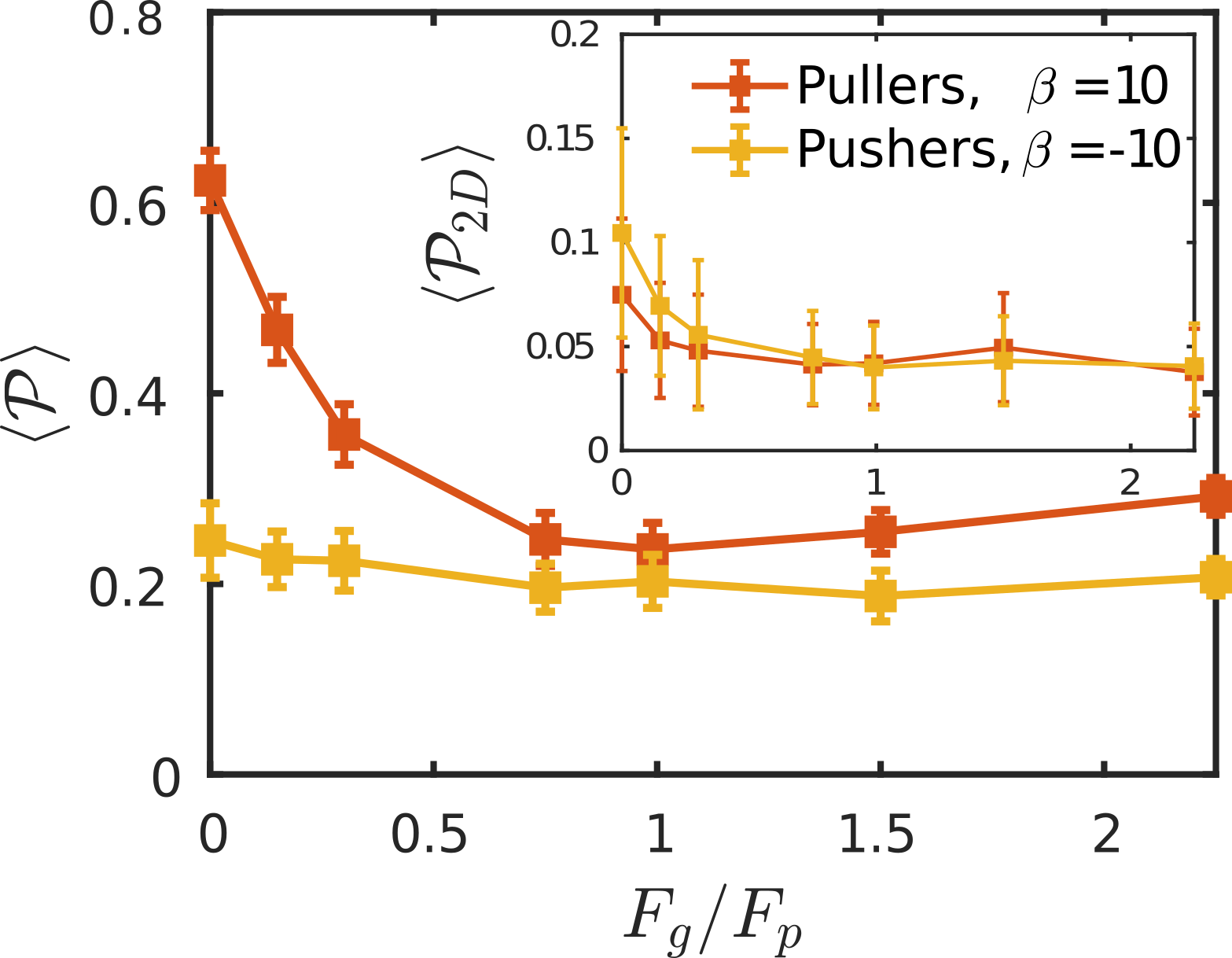}
 \caption{The average polar order parameter, $\langle \mathcal{P} \rangle$, as a function of the imposed gravitational field, $F_g/F_p$, for pullers and pushers at the bottom layer. \barri{Inset: the average polar order parameter computed with the projection of the orientation onto the $xy$ plane parallel to the wall $\langle \mathcal{P}_{\text{2D}} \rangle$}.}
 \label{fgr:GlobalPolarOP}
\end{figure}

\textcolor{black}{Having studied the probability distribution of the local polar order parameter, we now report  the global polar order parameter, $\langle \mathcal{P}\rangle$ in fig.~\ref{fgr:GlobalPolarOP} for pullers and pushers at the bottom layer, as a function of gravity.}

Although local alignment \textcolor{black}{as a  function of gravity} is similar for pullers and pushers,
\textcolor{black}{ $\langle \mathcal{P} \rangle$ shows a qualitatively different behavior.} 
\textcolor{black}{For pullers, it decreases as gravity increases, which can be attributed to the  hindrance of polar order with packing fraction~\cite{evans_orientational_2011},} \Barri{and the more frequent pairwise interactions.} The observed modest increase for $F_g/F_p>1$ might be due to the asphericity of the raspberry structure that might promote certain angles when  steric interactions with the bottom wall become stronger (see fig. 7 in SM).
There are two contributions to the global polar order at the bottom layer, the direct interactions among swimmers and indirectly, their alignment with the wall. \Barri{The inset of fig. \ref{fgr:GlobalPolarOP} shows that although the interaction among squirmers produces some polarization, the dominant contribution is the alignment with the wall. Similar to what happened previously with the local alignment, for $F_g/F_p<0.75$, the 2D polarization $\langle \mathcal{P}_{\text{2D}} \rangle$ is higher for pushers than for pullers, contrary to what happens in the 3D case. This is so since pushers swim side-by-side more stably than pullers (and thus contribute more to $\langle \mathcal{P}_{\text{2D}} \rangle$) and pullers have a more stable orientation when facing the wall than pushers (thus contributing more to $\langle \mathcal{P} \rangle$).  Although both $P\left(\mathcal{P}_{l,\text{2D}}\right)$ and $P\left(\mathcal{P}_l\right)$ are of the same magnitude, the same is not true for $\langle \mathcal{P}_{\text{2D}} \rangle$ and $\langle \mathcal{P} \rangle$, since the alignment with the wall contributes both to local and global polarizations, whereas the direct squirmer interactions contribute less to the global than to the local polarization (as different local polarizations cancel between each other).}

\begin{figure}[h!]
 \centering
 \includegraphics[width=8.9cm]{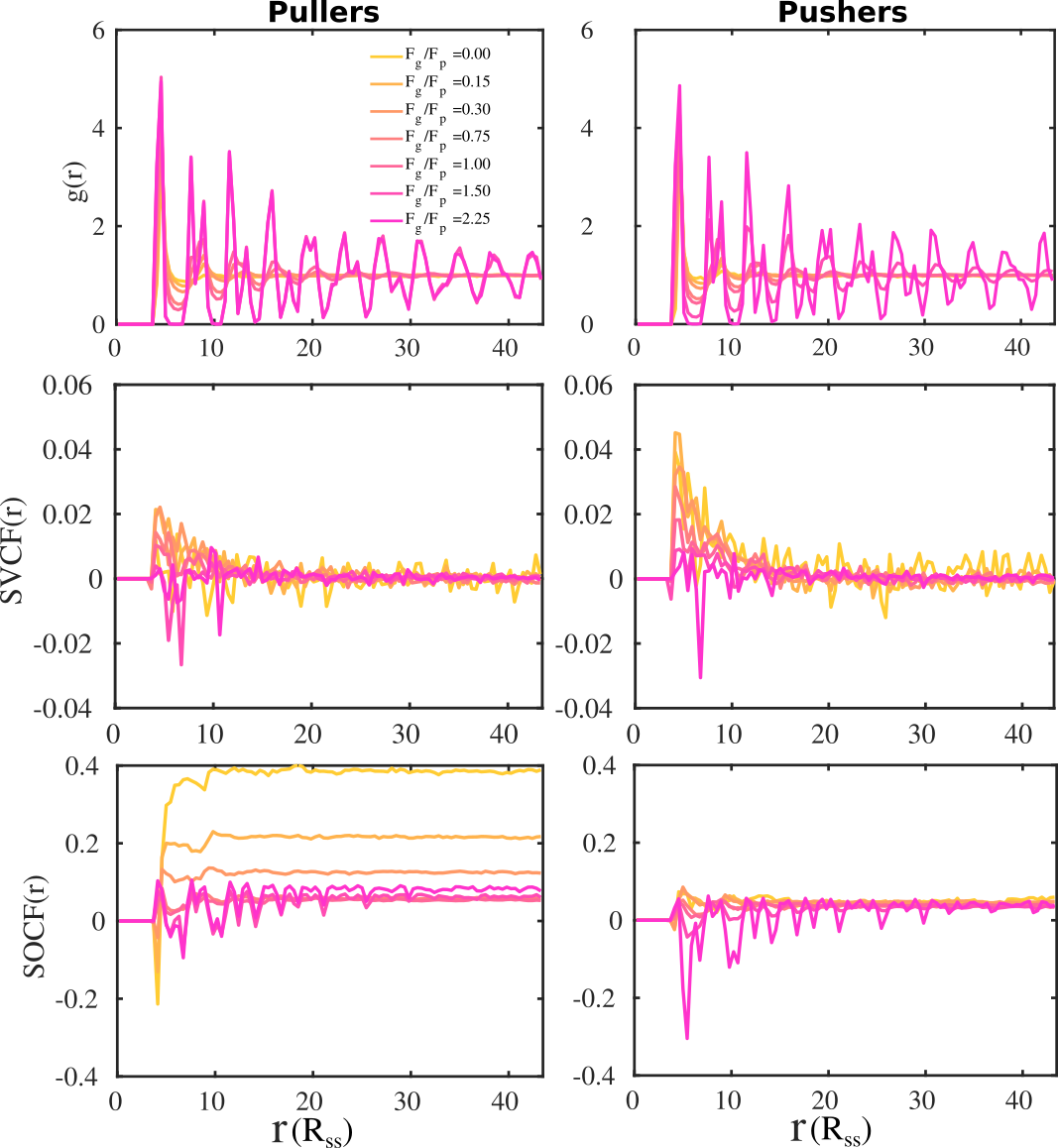}
 \caption{ Spatial correlations for pullers and pushers at different values of $F_g/F_p$ measured at the bottom layer. The distance is expressed in units of the solvent DPD cut-off , $R_{ss}$, and the bin size used to calculate the correlations is $0.435R_{ss}$.The pair correlation function, $g(r)$, is presented at the top row, the spatial velocity correlation function, $SVCF(r)$, in the central row, and the spatial orientation correlation function, $SOCF(r)$, in the bottom row.
 }
 \label{fgr:Correlations}
\end{figure}


\barri{Regarding $\langle \mathcal{P}_{\text{2D}} \rangle$ similar, albeit larger, polar order has been found in ref. \citenum{kuhr2019collective} in which swarming is found for pushers but not for pullers.}
\barri{Regarding $\langle \mathcal{P}\rangle$}, although 
global polar order has \barri{also} been found in previous studies in  squirmer suspensions either in bulk\cite{ishikawa_development_2008,evans_orientational_2011,alarcon2013spontaneous}, in confinement\cite{oyama_purely_2016} and under gravity\cite{enculescu_active_2011}, 
in our system, \barri{the behavior of $\langle \mathcal{P}\rangle$ is mostly controlled by} 
the interplay between the hydrodynamic coupling to the wall and the fluctuations induced by temperature\cite{zottl_hydrodynamics_2014} and stresslets\cite{alarcon2013spontaneous}.
The increased puller polarization  induced by the  wall \barri{($\langle \mathcal{P}\rangle$ in fig. \ref{fgr:GlobalPolarOP}) correlates with their aggregation at the bottom layer ($\phi_{\text{2D}}$ in fig. 4 of SM) for low gravitational fields \cite{ruhle2018gravity}.} 

\barri{Although previous studies have argued that asymptotic near-field hydrodynamics strongly determines squirmer interactions and alignment\cite{gotze_mesoscale_2010,ishikawa_development_2008, delfau2016collective, ruhle2018gravity}, in our case we see some global polar order in $\langle \mathcal{P}_{\text{2D}} \rangle$ without the need of lubrication corrections. In ref. \citenum{alarcon2013spontaneous} large global polar order is also found for bulk suspensions of pullers also without lubrication corrections. }
In order to analyze in more detail the alignment of squirmers with the wall, 
we calculate the distribution, $P(\alpha)$, of the angle between the colloid orientation and the normal vector to the bottom wall (fig. \ref{fgr:Distributions}, bottom row)\textcolor{black}{; hence $\alpha = 0^o$ corresponds to the squirmer pointing upwards, $\alpha = 180^o$ pointing downwards, and $\alpha = 90^o$  } oriented parallel to the wall in the $xy$ plane.

\textcolor{black}{In the absence of gravity, the distribution for pullers}
displays two peaks, \textcolor{black}{as they tend to align with}
the wall forming angles $\alpha \approx 125^o$ and $155^o$. 
\Barri{Additionaly, a valley around $\alpha = 90^\circ$ shows the low probability of pullers being oriented parallel to the wall}. For pushers \Barri{the distribution takes non-negligible values around $\alpha = 90^\circ$ as expected, and we also note how}
\textcolor{black}{a peak develops at $\alpha \approx 115^o$ while the one at $155^o$ is absent}. \Barri{This difference in the wall alignment  both for  pullers and pushers has also been  observed in LB simulations \cite{ouyang2018hydrodynamic}.} Not surprisingly, \Barri{the most probable} angles are far from the theoretical prediction for a single squirmer\cite{lauga_hydrodynamics_2009}.
In ref. \citenum{shen2018hydrodynamic} the authors find tilt angles of $\alpha=132^{\circ}$ for pullers and $\alpha=62^{\circ}$ for pushers of strength $\beta=\pm 9.6$ in LB simulations of a single squirmer interacting with a wall. Our work is  in  a qualitative agreement for pullers, although the angles for pushers are in sharp contrast. Again, this is not surprising since in our case interactions between squirmers, thermal fluctuations and asphericity of the swimmers need to be taken into account. In ref. \citenum{li_hydrodynamic_wall_2014} the authors carry out computational fluid dynamics simulations (CFD) for a single and multiple squirmers between walls and find that although a single squirmer with $\beta < 1$ is scattered by the wall with $\alpha < 90^{\circ}$ (in contrast to ref. \citenum{shen2018hydrodynamic} where squirmers scatter for $|\beta|<2$), when multiple squirmers are present, they  aggregate near the wall and their orientation is tilted towards it with $\left\langle\alpha\right\rangle\approx 135^{\circ}$ (for $\beta = 3$ ) and $\approx 120^{\circ}$ (for $\beta=-3$), which is in better agreement with our results.

\Barri{As gravity increases, the probability of finding a pusher parallel to the wall decreases, reaching values close to zero for the highest gravitational fields applied. Unlike pullers, pushers are rarely oriented towards the wall forming angles around $\alpha = 155^o$. Instead, when pointing towards the wall, pushers form a preferred angle around $\alpha = 125^o$, as can be observed from the peak in the distributions. Such a peak becomes more pronounced as the gravitational field increases. Similar to pullers, when pointing away from the wall pushers also form angles around $\alpha = 60^o$, 
but in this case the peak in the distributions is present for all  gravitational fields considered. These peaks can be seen in the bottom left panel of fig. \ref{fig:SedimentationProfiles} (see  fig. 3 of SM for a zoom):  the two first peaks  correspond to the negative peak $\left\langle\cos\alpha\right\rangle = -0.7$, that shrinks to $\left\langle\cos\alpha\right\rangle = -0.5$ when increasing gravity, and the third one to the positive peak $\left\langle\cos\alpha\right\rangle = 0.5$ at slightly higher $z$. As displayed in fig. 3 of SM, the variation of $\left\langle\cos\alpha\right\rangle$ with height is smooth for pullers, even reaching a minimum at $z\approx 0.95 D_{RDF}$, and sharper for pushers,  departing from a non-zero derivative and increasing almost linearly without passing through any minimum.}



\Barri{At the highest gravitational fields, the $P(\alpha)$ distributions for pullers and pushers are very similar, suggesting that hydrodynamics plays a \textcolor{black}{smaller}
role in the orientation of microswimmers with respect to the wall. 
The distributions feature a short peak around $\alpha = 60^o$ and a high peak around $\alpha = 125^o$, indicating that although the preferred orientation of the squirmers is still pointing towards the wall, \textcolor{black}{some point away, unlike  orientations  observed at low gravity.}
This feature can also be detected in fig. 3 of the SM where the peak in $\phi_{\text{2D}}$ happens at heights where $\left\langle\cos\alpha\right\rangle$ is negative, while the maximum values of $\left\langle\cos\alpha\right\rangle$ correspond to a $\phi_{\text{2D}}$ an order of magnitude lower.} The preferred angles at these high gravitational fields are also related to the swimmers' raspberry structures. 
Although the DPD cut-off radius makes the swimmers approximately  spherical, there is still an steric contribution to the settled positions of the colloids on the bottom wall (see fig. 7 of SM). 

\Barri{For completeness, we briefly review the possible mechanisms contributing to the vertical squirmer orientation.} In a recent simulation work on ellipsoidal squirmers confined in thin layers, angles around $60^o$ and $120^o$ with respect to the normal axis of the wall were also observed\cite{wu2024collective}. For purely spherical squirmers, a more pronounced upward orientation is seen in ref. \citenum{kuhr2019collective} that is the highest for neutral squirmers and decreases with increasing $|\beta|$, the authors also find that pushers tilt away from the vertical and their height increases as, they argue, they move through a ``transitional region between the upright orientation at the wall as calculated in lubrication theory and the parallel far-field orientation''. The effect of the interplay between far-field and lubrication in the swimmers orientation is also analyzed in refs. \citenum{schaar_detention_2015} and \citenum{thery_lauga_maas_2023}. 
To gain further information on the alignment of the microswimmers at the bottom layer, we compute the SVCF and SOCF.
In fig. \ref{fgr:Correlations} we present the pair correlation function, $g(r)$ (top panels),  the $SVCF(r)$ (middle panels)  and the $SOCF(r)$ (bottom panels). 


As already noted from the SSF (See fig. \ref{fgr:SSF2D}), the $g(r)$ (top panels) show how  swimmers at the bottom layer form a hexagonal array as the gravitational field increases for both suspensions of pullers and pushers. The $SVCF(r)$ (middle panels) for both pullers and pushers show a positive correlation at short distances that rapidly decays for $F_g/F_p <= 1.0$. Upon increasing gravity, the correlation for pushers monotonically decreases 
to negative values at the highest imposed gravitational field. However, for pullers the behavior is non-monotonic, first increasing up to $F_g/F_p = 0.15$ and then decreasing, assuming negative values for $F_g/F_p >=1.50$. 

It is noteworthy that the $g(r)$ (top panels) and the $SOCF(r)$ (bottom panels) are correlated, especially at high gravitational fields, indicating that positive and negative correlations of the swimmer orientation are somehow correlated at the same time with the hexagonal structure of the bottom layer. 
For low gravitational fields ($F_g/F_p <=0.30$), the first peak of the $SOCF(r)$ \textcolor{black}{for pullers}  is negative, being less pronounced as the gravitational field decreases. This means that at short-range  pullers tend to orient in a partially anti-parallel way with each other. At longer distances, we observe how the contributions of the $SOCF(r)$ becomes positive (squirmers are oriented partially parallel) and then reach a constant value, being the case with no gravity the one  corresponding to the highest positive correlation. For the highest $F_g/F_p$, the $SOCF(r)$ reveals that  pullers orientation positively contribute  to the correlation at short-range. Then, at longer distances, the $SOCF(r)$ describes the hexagonal structure of the layer in such a way the peaks of the $g(r)$ coincide with the regions of positive correlation, while the minima of the $g(r)$ with regions of negative correlation. Finally, the $SOCF(r)$ of pusher suspensions  also correlate well with the hexagonal structure of the bottom layer at high gravitational fields, with the difference that now the regions of negative correlations are more pronounced than those of the $SOCF(r)$ for pullers. Additionally, at low gravitational fields, the $SOCF(r)$ for pushers acquires small positive values at long-range.

The results shown for the $SOCF(r)$ (bottom panels in fig.\ref{fgr:Correlations}) match with the results of $P(\mathcal{P}_l)$ and $P(\alpha)$ presented in fig. \ref{fgr:Distributions}. The highest positive correlations are obtained for suspensions of pullers at low gravitational fields, where they are mainly oriented towards the wall; whereas suspensions of pushers exhibit rather small positive correlations at long-range, since their local alignment is low and their orientation with respect to the wall is distributed over a wider range of angles. 
The $SOCF(r)$ of pullers can also be correlated with the global polar order $\langle \mathcal{P} \rangle$ (fig. \ref{fgr:GlobalPolarOP}). 
Although the $SOCF(r)$ displays a constant value spawning the full simulation box,
computing it for the projected orientations onto the $xy$ plane parallel to the wall, $SOCF_{\text{2D}}(r)$ (SM fig. 6), shows that the correlations \textcolor{black}{vanish},
suggesting that the observed correlations are due to the vertical alignment of squirmers and are \textcolor{black}{modulated by squirmer  interactions, whose relevance increases with gravity.}

\section{Conclusions}

\barri{In summary, 
DPD simulations \Barri{have been performed} to study  sedimenting raspberry-like colloidal suspensions bounded between parallel walls (fig. \ref{fgr:Colloid}). When  colloids are passive, the observed sedimentation velocity (fig. \ref{fig:sedimentation}) is \Barri{qualitatively} well captured by Stoke's law \Barri{considering a height-independent friction coefficient}, with a small overestimation ascribable to the packing fraction correction for collective sedimentation\cite{Russel_1989_colloidal_dispersions} \Barri{and to the varying friction coefficient induced by the walls\cite{ruhle2018gravity}.}
%
\textcolor{black}{When passive colloids are replaced by squirmers, in the absence of gravity they accumulate  at walls due to activity, as expected} (e.g. fig. \ref{fig:SedimentationProfiles}).
\Barri{This aggregation is} greater for pullers than for pushers due to the interplay between hydrodynamics and temperature, agreeing with previous studies\cite{li_hydrodynamic_wall_2014,ishimoto_squirmer_boundary_2013,schaar_detention_2015} \Barri{(although other works report different results\cite{llopis_hydrodynamic_wall_2010,shen2018hydrodynamic}).} 
Near the walls, the orientation for both pullers and pushers is tilted to the wall,  more prominently for pullers than for pushers, consistently with far-field hydrodynamics\cite{lauga_hydrodynamics_2009}. 
We noted that the characteristic \Barri{orientation} angles observed 
are also influenced by squirmer interactions, the absence of lubrication corrections and the asphericity of the colloid. Far from the walls, the swimmers vertical orientations distribute isotropically. 
Consistently with previous studies\cite{kuhr2017collective,scagliarini2022hydrodynamic}, at moderate gravitational fields, where swimmers  can still reach the top wall, we observe an exponential decay of the squirmer packing fraction (fig. \ref{fig:SedimentationProfiles}) with sedimentation lengths that are always higher than for passive colloids, being greater for pushers than for pullers and decreasing with increasing gravity. We argued that pushers are able to access higher regions due to pairwise far-field hydrodynamic interactions near a boundary described in a recent work\cite{thery_lauga_maas_2023}. We observed no sign of previously reported bioconvenction\cite{kuhr2017collective} and argued that the geometry of the simulation domain and the strength of the stresslet may be responsible. However, polar order is observed in the bulk region where swimmers orientations tilt upwards as we increase the gravitational field. 
%
As gravity is increased further, 
\Barri{certain} heights become inaccessible to swimmers, the exponential regime is broken and two well defined bottom layers begin to develop.}


\textcolor{black}{Once colloids have settled}, 
the structure of the bottom layer \Barri{was analyzed}. \Barri{It was} 
observed that the bottom layer undergoes a transition from an isotropic liquid to a hexagonal crystal as gravity is increased (see fig. \ref{fgr:SSF2D}, fig. \ref{fgr:Distributions}, first row, and fig. \ref{fgr:G6}). In the reference case of passive colloids, 
the hexagonal crystal contains many defects, due to the fact that the system is kinetically trapped in states that  thermal motion is not able to overcome.  For  pullers and pushers  \Barri{there was no observation of} 
any kinetically trapped state, since their activity allows their rearrangement into more ordered structures. However,  pullers tend to preserve the hexagonal order better than pushers: this  is reflected in the fact that a perfect hexagonal crystal is obtained for pullers at lower gravitational fields. This is related to the hydrodynamic field generated by pullers and pushers. For the former,  hydrodynamics favours the orientation of the swimmers pointing towards the wall with a certain persistence that hinders their mobility; whereas in the latter it favours orientations of the swimmers parallel to the wall, making them more motile and thus more prone to alter the order of the bottom layer. 

\Barri{It was} 
also observed that, although the local polar alignment is similar in both pullers and pushers despite the gravitational field (See fig. \ref{fgr:Distributions}, central row), the global polar alignment is higher for pullers than for pushers for all $F_g/F_p$ and shows a non-monotonic behaviour with a minimum at $F_g/F_p = 1.0$.  While pushers always display a low global polar alignment (See fig. \ref{fgr:GlobalPolarOP}).
%
In the absence of gravity, the dominant contribution to the high polar order exhibited by pullers comes from their interactions with the wall and is modulated by squirmer pairwise interactions.
The reduction in this polar order, when gravity is increased, was explained taking into account the increase of the packing fraction at the bottom layer, which has been shown to hinder polar order.
The difference in polar order between pullers and pushers was explained through the interplay of hydrodynamics and thermal fluctuations, being pullers more stable under reorientations.
%

\Barri{It was} found that at high gravitational fields,  both pullers and pushers not only are mainly oriented towards the wall, forming an angle with respect to the wall's normal vector of about $120^o$, but are also oriented pointing away from the wall with a preferred angle of about $60^o$ (See fig. \ref{fgr:Distributions}, bottom row). As described above, this preferred orientations with respect to the wall do not correlate with the polar alignment of the microswimmers at high gravitational fields. However, from low to moderate gravitational fields, 
pushers tend to align parallel to the wall (high values of $P(\alpha)$ around $\alpha = 90^o$) whereas pullers tend to align towards the wall forming angles of $125^o$ and $155^o$, which explains the high polar order exhibited by pullers and the low polar order of pushers. 

This pictures becomes clearer when analysing the spatial correlation functions at the bottom layer (see fig. \ref{fgr:Correlations}). Apart from the transition from an isotropic liquid to a hexagonal crystal also observed from the pair correlation functions (see fig. \ref{fgr:Correlations}, top row), we confirm that the orientations of the pullers are correlated throughout the entire system from low to intermediate gravitational fields, whereas pushers display low correlation (see fig. \ref{fgr:Correlations}, bottom row). Interestingly, at high gravitational fields, although the $SOCF(r)$ are rather low at long distances for pullers and pushers, at short distances they correlate with the $g(r)$, showing positive correlations at the $g(r)$ maxima and negative correlations at the $g(r)$ minima, being this effect more pronounced for pushers. Regarding the $SVCF(r)$, the behaviour is very similar for both microswimmer types, showing positive correlations for low to intermediate gravitational fields at short distances, and negative correlations at short distances for high gravitational fields. 

To conclude, there are many possible extensions with the possibility of practical applications of the presented study. Few possible studies 
to develop in the coming future include: an analysis of the dynamics of the deposited bottom layer, the study of the effects of including lubrication corrections, broader scans of parameters such as propulsion force, squirmer parameter or temperature to promote pair-wise squirmer alignment, sedimentation of mixtures of active and passive colloids and different colloid and/or bounding geometry.

\section*{Conflicts of interest}
There are no conflicts to declare.

\section*{Data Availability Statement}
The data that support the findings of this study are available from the corresponding author upon reasonable request.

\section*{Supplementary Material}

The present manuscript is accompanied by a document \verb$supp_info.pdf$ with additional results and 4 video files:
\begin{itemize}
    \item \verb$video1.avi$: realistic animations of 9 of the studied simulations showing the time evolution of passive, puller and pusher systems under null, moderate and strong gravitational force. The orientation of the colloids is displayed as a red bead in each of the raspberry colloids.
    \item \verb$video2_psi6.avi$: the video file contains an animated version of fig. \ref{fgr:SSF2D} without the SSF insets.
    \item \verb$video3_cos-alpha.avi$: same as \verb$video2_psi6.avi$ but the color code shows the vertical orientation of swimmers measured as $\cos\alpha$.
    \item \verb$video4_local-pol.avi$: same as \verb$video2_psi6.avi$ but the color code shows the local polarization.
\end{itemize}

\section*{Acknowledgements}

C.V. acknowledges fundings  IHRC22/00002 and 
PID2022-140407NB-C21 from MINECO. This project has received funding from the European Union’s Horizon research and innovation programme under the Marie Skłodowska-Curie grant agreement No 101108868 (BIOMICAR).  I.P. acknowledges  financial support to DURSI under Project No. 2021SGR-673, Ministerio de Ciencia, Innovaci\'on y Universidades MCIU/AEI/FEDER under
grant agreement PID2021-126570NB-100 AEI/FEDER-EU.and  Generalitat de Catalunya for financial support under Program Icrea Acad\`emia. We acknowledge MARENOSTRUM-BSC (grant FI-2024-2-0044). C.M.B.G acknowledges enrichful discussions with José Martín-Roca, Juan Pablo Miranda López and Rodrigo Fernandez-Quevedo. 

\balance

\bibliography{GW}
\bibliographystyle{ieeetr}


\end{document}


\maketitle
\begin{figure}[h!]
    \centering
    \includegraphics[width=0.33\textwidth]{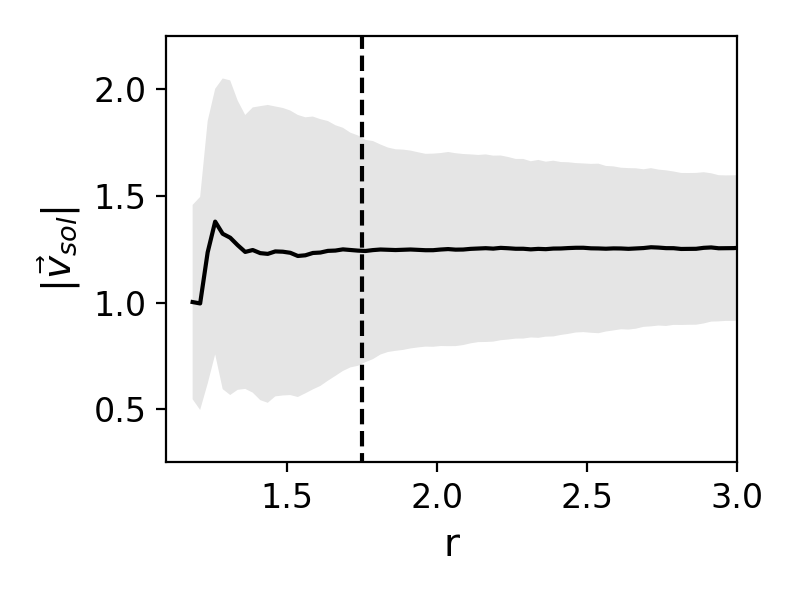}    \includegraphics[width=0.415\textwidth]{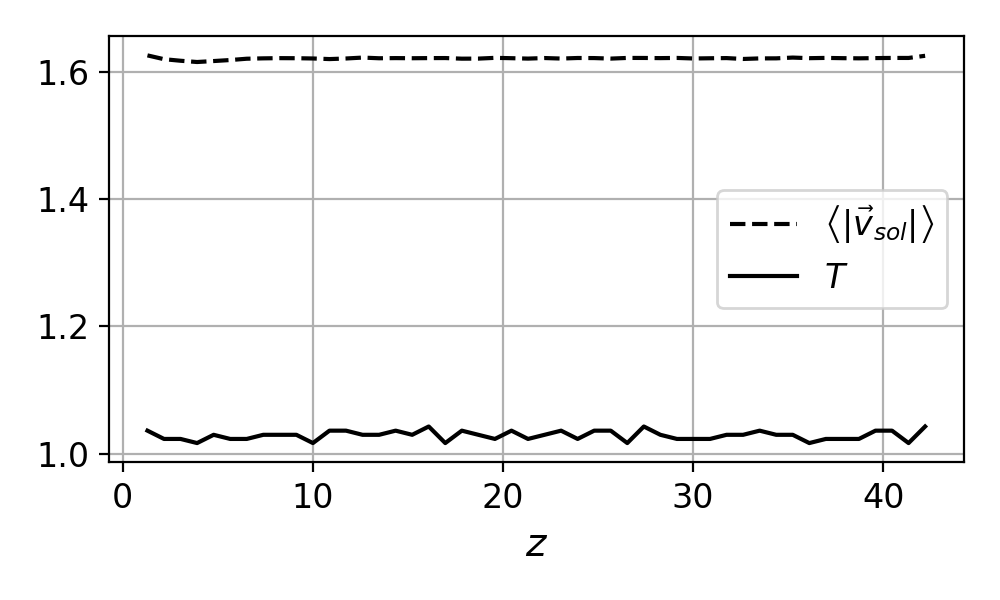}
    \caption{\textbf{Left}: Passive colloids boundary condition in bulk without gravity. \textbf{Right}: Temperature profile along the vertical and average norm of the solvent velocity in the passive colloidal dispersion under gravity $F_g=34$.}
    \label{fig:enter-label}
\end{figure}

\begin{figure}[h!]
    \centering
    \includegraphics[width=0.34\textwidth]{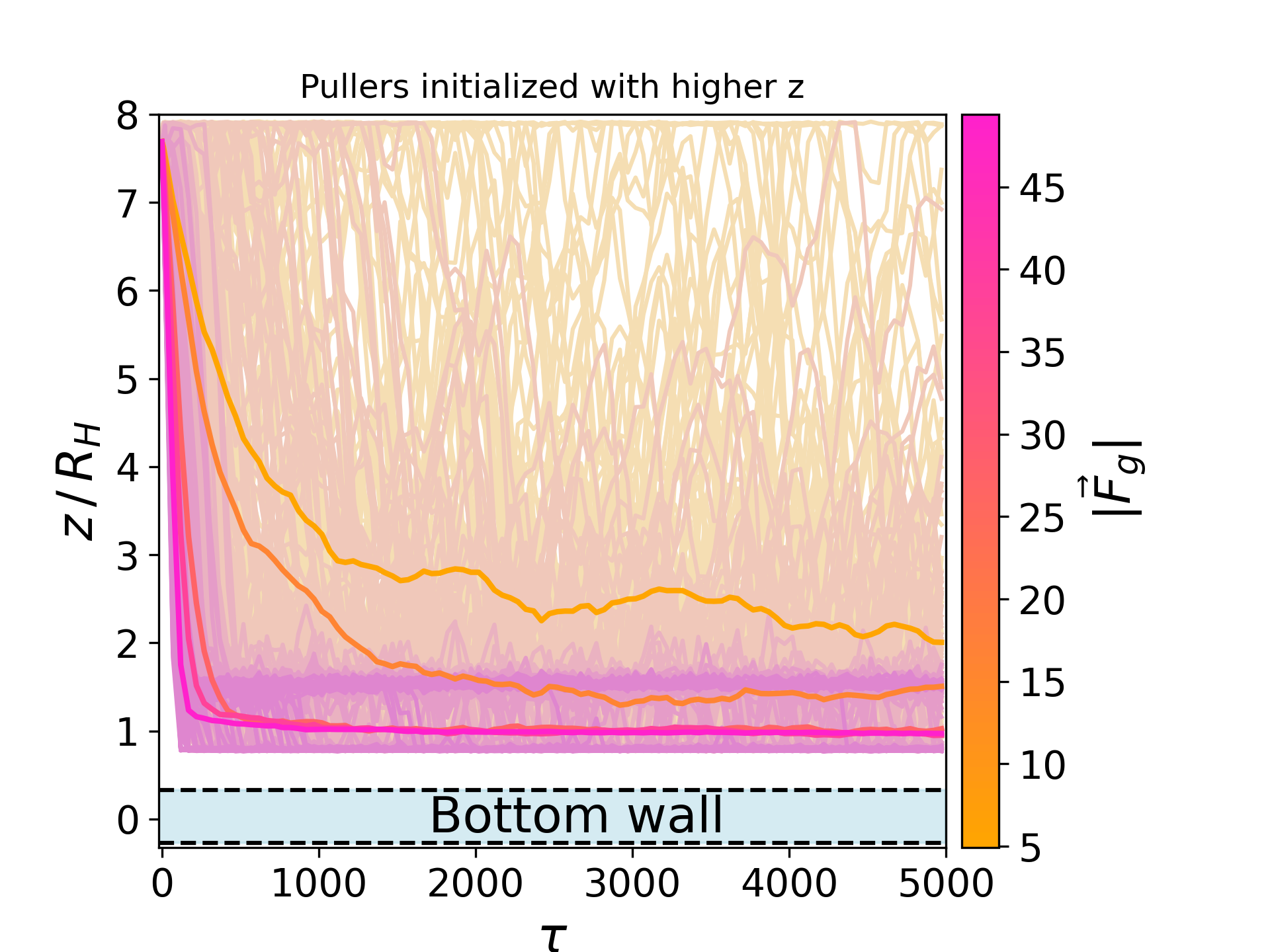}\includegraphics[width=0.34\textwidth]{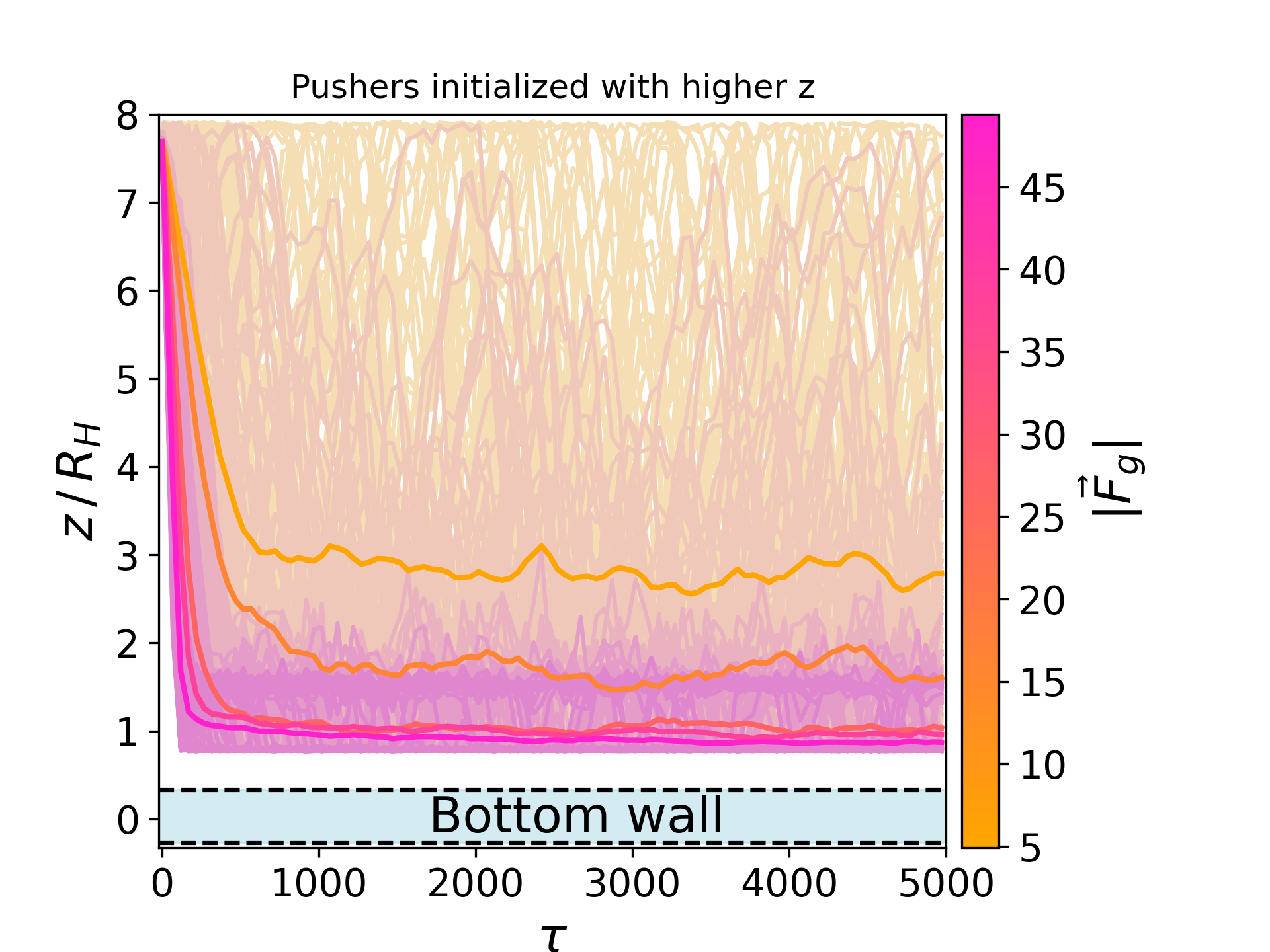}\includegraphics[width=0.31\textwidth]{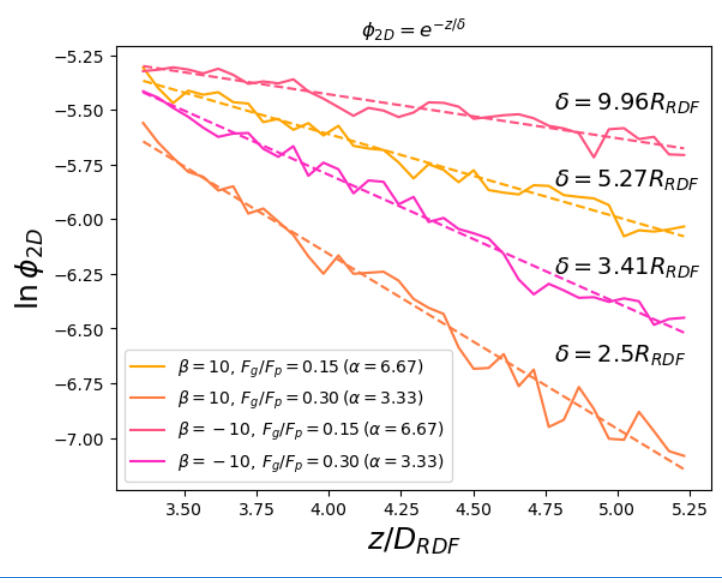}
    \caption{\textbf{Left and center}: Sedimentation trajectories (light colors) and their average (dark colors) for the 100 topmost initialized colloids. \textbf{Right}: Sedimentation lengths, $\delta$, computed by fitting sedimentation profile in the shown range to $\phi_{2D} = e^{\,z/\delta}$. In the legend $\alpha$ is just the inverse of $F_g/F_p$ to compare with \cite{kuhr2017collective}.}
    \label{fig:enter-label}
\end{figure}

\begin{figure}[h!]
    \centering
    \includegraphics[width=0.66\textwidth]{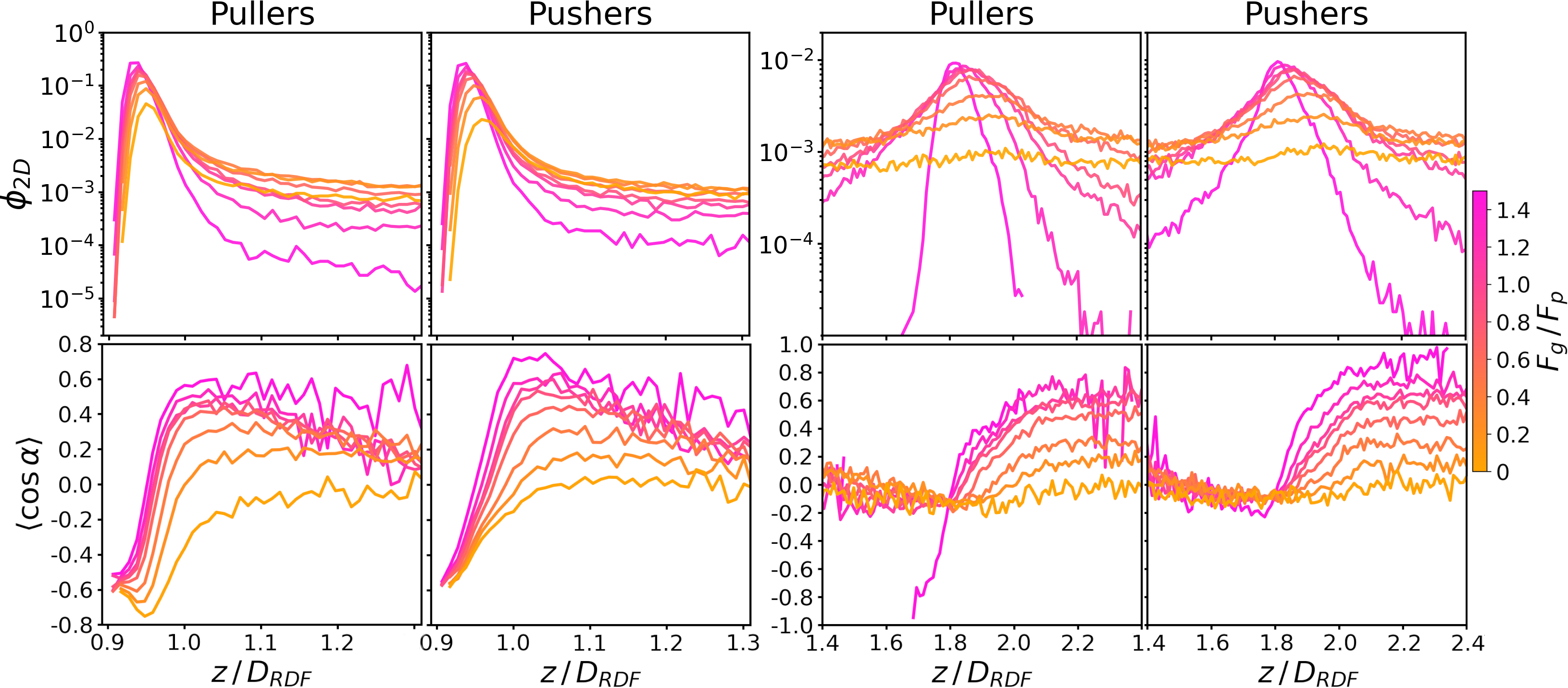}
    \caption{\textbf{Left}: Zoom on first layer using a bin width of $\Delta z = 0.044$. \textbf{Right}: Zoom on second layer using a bin width of $\Delta z = 0.044$.}
    \label{fig:enter-label}
\end{figure}

\begin{figure}[h!]
    \centering
    \includegraphics[width=0.33\textwidth]{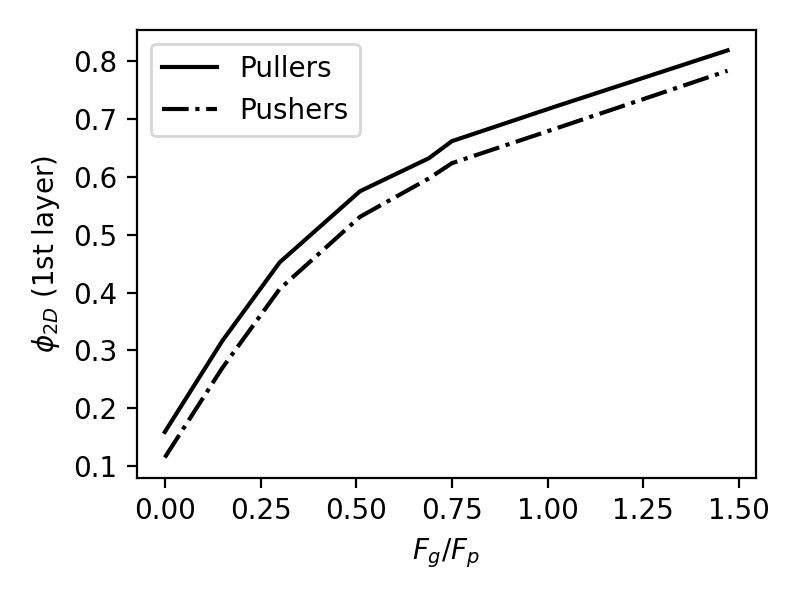}\includegraphics[width=0.33\textwidth]{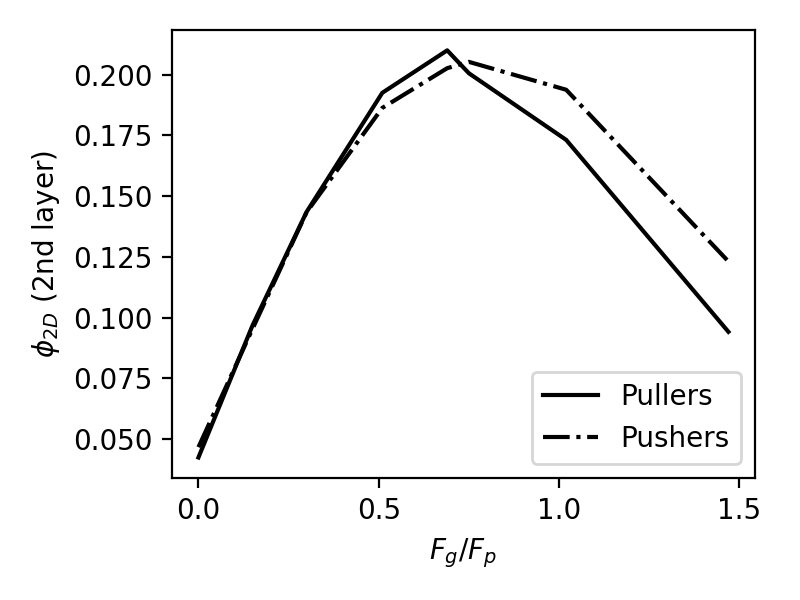}\includegraphics[width=0.33\textwidth]{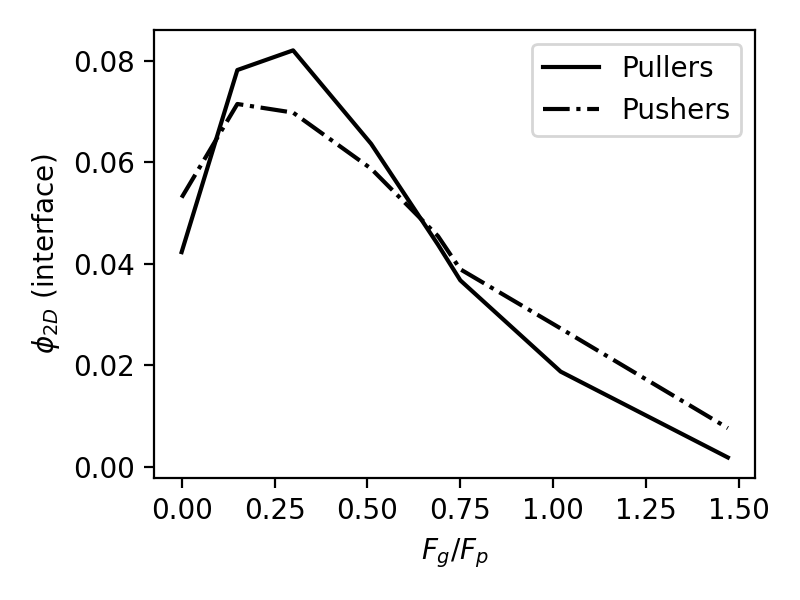}
    \caption{Maxima of the two dimensional packing fraction for 1st layer (left), the 2nd layer (center) and the interface between both (right). These plots were obtained using a bin width equal to the colloids radius estimated via RDF, $\Delta z = R_{RDF}\approx 2.1$, }
    \label{fig:enter-label}
\end{figure}

\begin{figure}[h!]
    \centering
    \includegraphics[width=\textwidth]{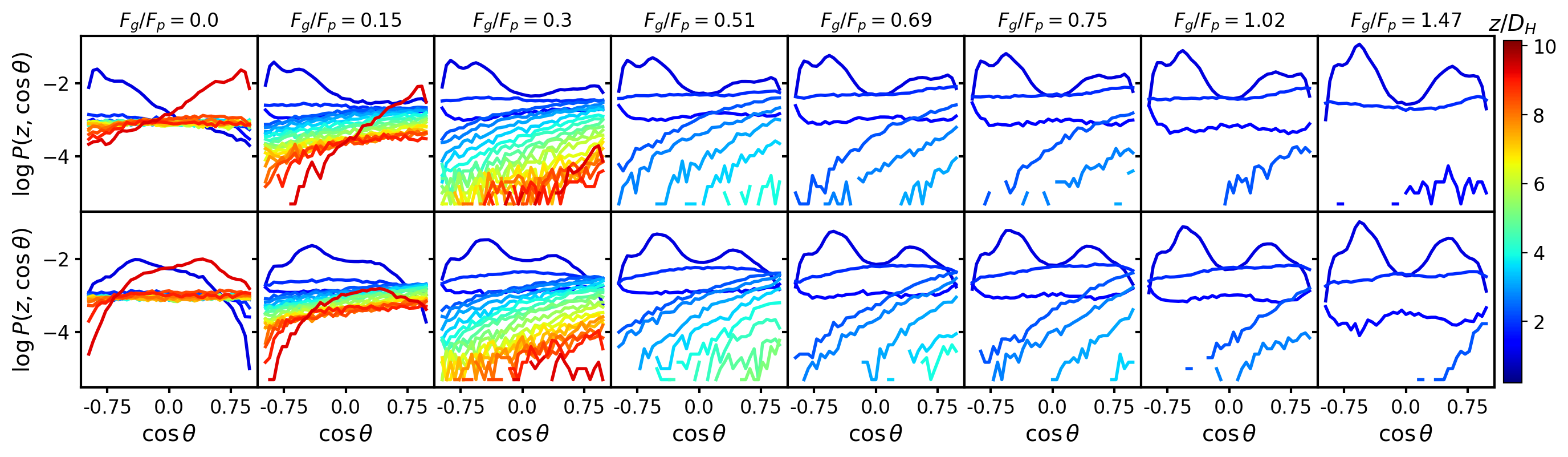}
    \caption{\textbf{Top row}: pullers. \textbf{Bottom row}: pushers. To be compared with fig. 6 of ref. [\citenum{scagliarini2022hydrodynamic}], fig. 11 of ref. [\citenum{li_hydrodynamic_wall_2014}], fig. 2 of ref. [\citenum{kuhr2019collective}] and fig. 2 of ref. [\citenum{enculescu_active_2011}].}
    \label{fig:enter-label}
\end{figure}


\begin{figure}[h!]
    \centering
    \includegraphics[width=\textwidth]{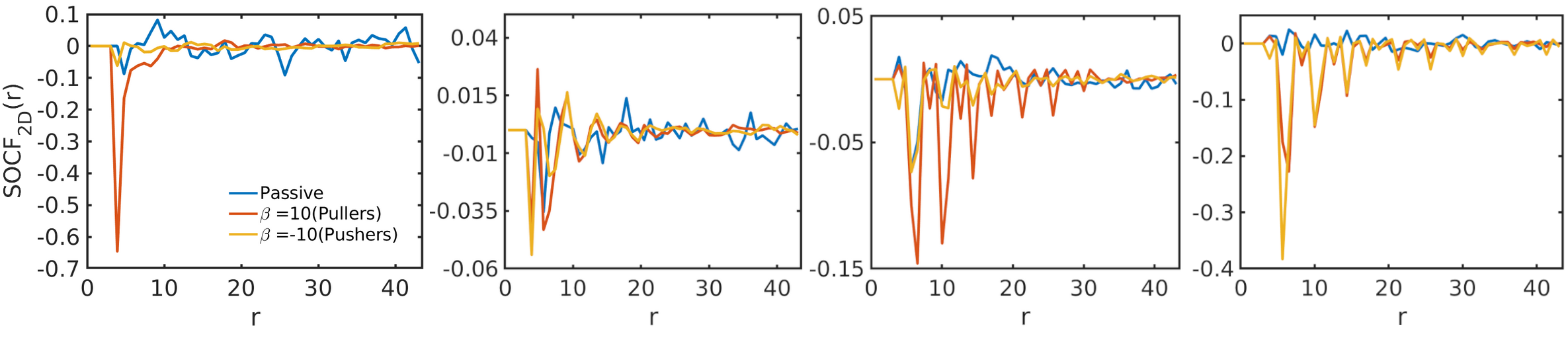}
    \includegraphics[width=\textwidth]{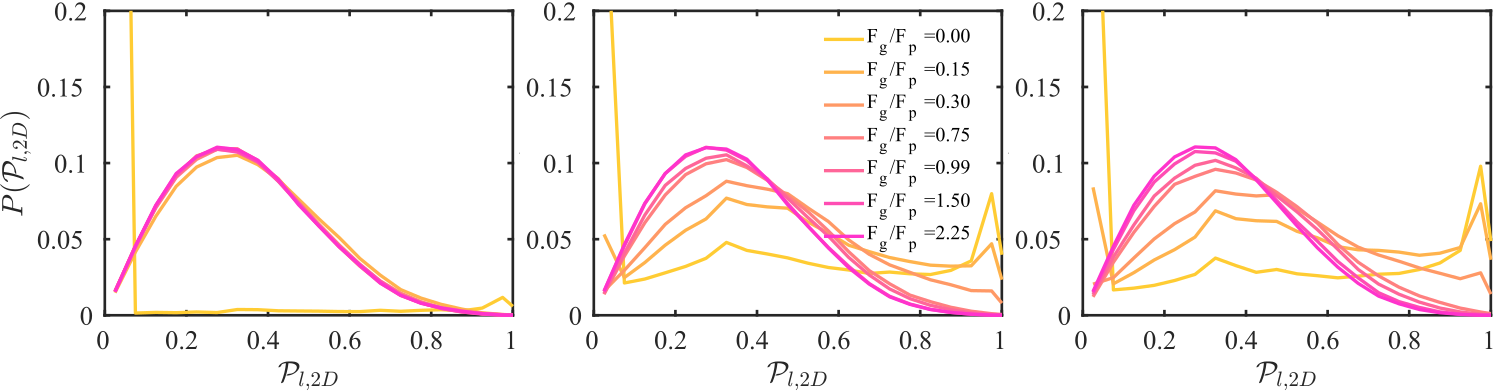}
    \caption{\textbf{Top:} The spatial orientation correlation functions computed with the projection of the orientation of the squirmers onto the $xy$ plane.  From left to right: $F_g/F_p = 0.0,\,0.75,\,1.5,\,2.25$. \textbf{Bottom:} Distributions for local polarization computed with the projection of the orientation of the squirmers onto the $xy$ plane for passive (left), pullers (center) and pushers (right).}
    \label{fig:enter-label}
\end{figure}

\begin{figure}[h!]
    \centering
    \includegraphics[width=\textwidth]{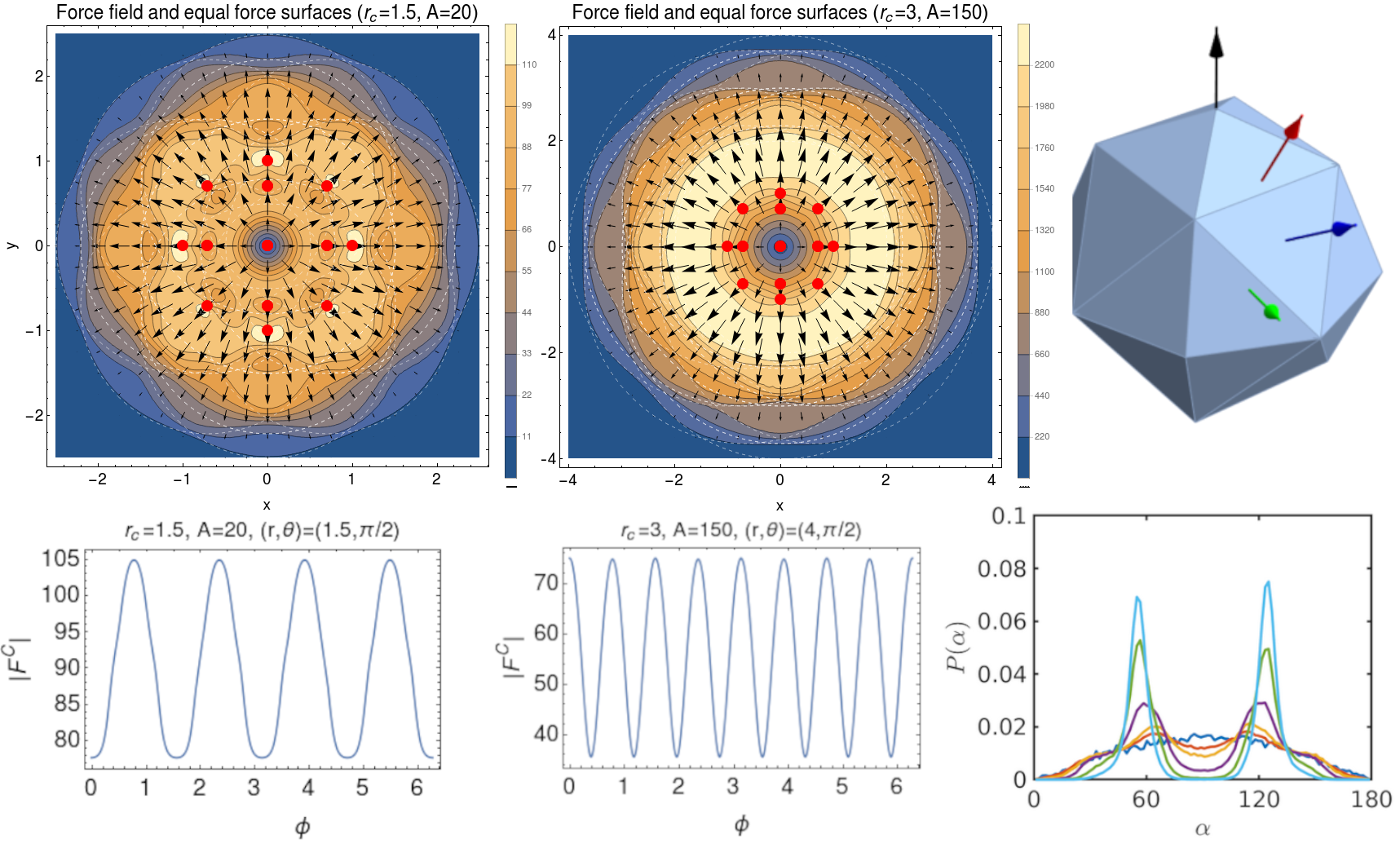}
    \caption{\textbf{Top:} Equatorial section of the force field generated by the conservative interaction of the raspberry colloid composed of 19 particles (red dots) for a DPD cutoff of $r_c=1.5$, $A=20$ (left) and $r_c=3$, $A=150$ (center). Polyhedral structure of the colloid (right), the filler particles are placed at the vertices of this polyhedron. The black arrow is the orientation of the colloid, the red, green and blue ones correspond to the normal direction of each polyhedron face, the rest of the faces are symmetric to one of these. The angles between the color arrows and the vertical (black arrow) are: 30.4º (red), 54.7º (blue) and 69.1º (green). The angles formed by the normals of the faces symmetric to the previous ones are obtained by substracting the angles to 180º: 149.6º (red), 125.3º (blue) and 110.9º (green). At first sight the most stable resting position of such a polyhedron under gravity is the face corresponding to the blue arrow. \textbf{Bottom:} Force along the equatorial coordinate $\phi$ at the $r$ and $\theta$ shown in the plots label for the previous cases (left and center). Distribution of orientation angle w.r.t. the wall normal for passive colloids for increasing gravitational field (as $F_g$ increases the two peaks start to develop).}
    \label{fig:enter-label}
\end{figure}

\begin{figure}[h!]
    \centering
    \includegraphics[width=\textwidth]{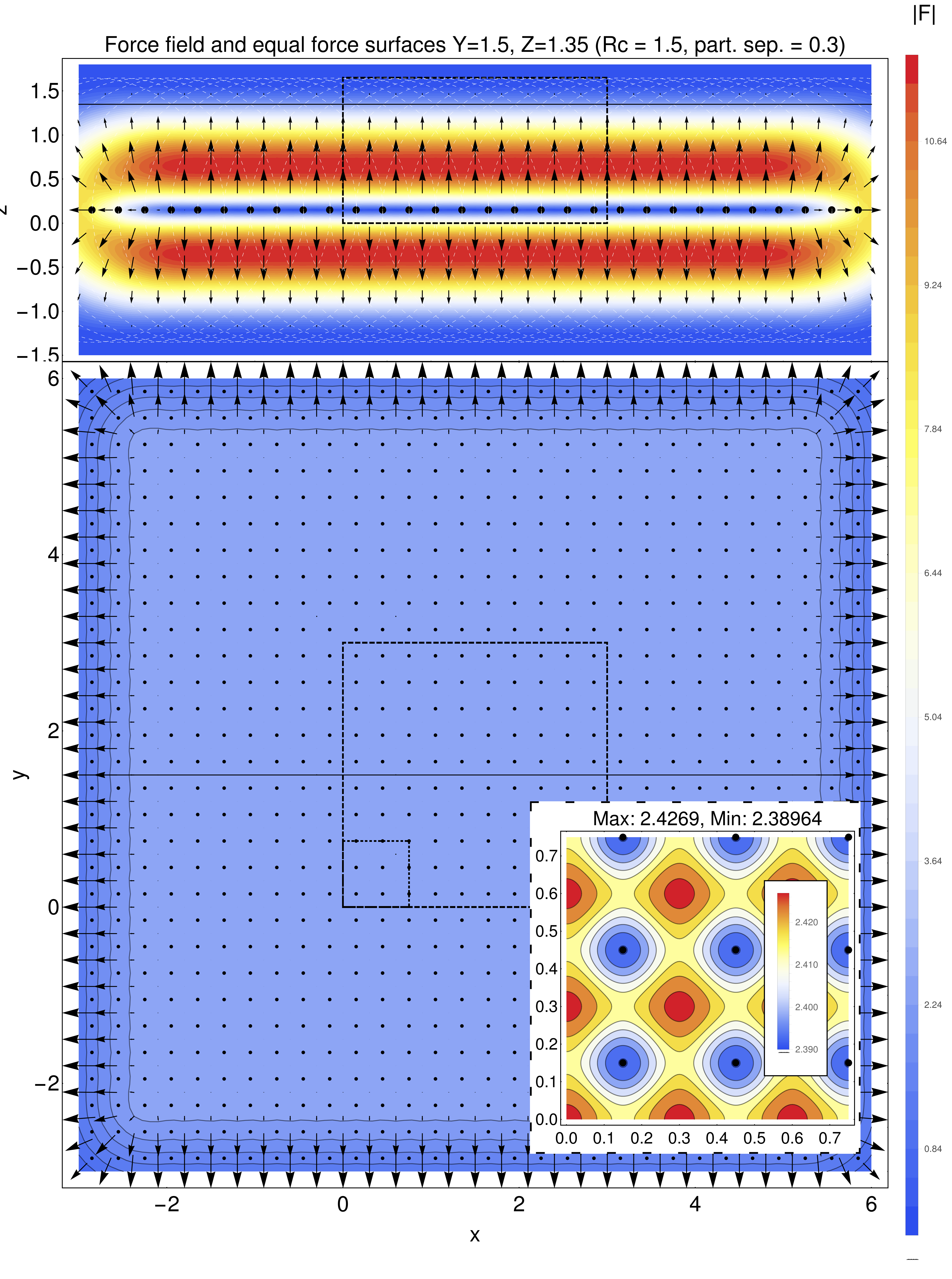}
    \caption{DPD conservative force field and equal force surfaces (color) for smaller a wall region (dashed rectangle) of 10 particles (black dots) by side with their 8 periodic images to the sides. \textbf{Top:} XZ section for $y=1.5$. An homogenous upward repulsive force is observed. \textbf{Bottom:} XY section at $z=1.35$. \textbf{Inset}: zoom on the lower left corner of the wall, showing wall rugosity and maximum and minimum values of the force.}
    \label{fig:enter-label}
\end{figure}

\begin{figure}[h!]
    \centering
    \includegraphics[width=0.65\textwidth]{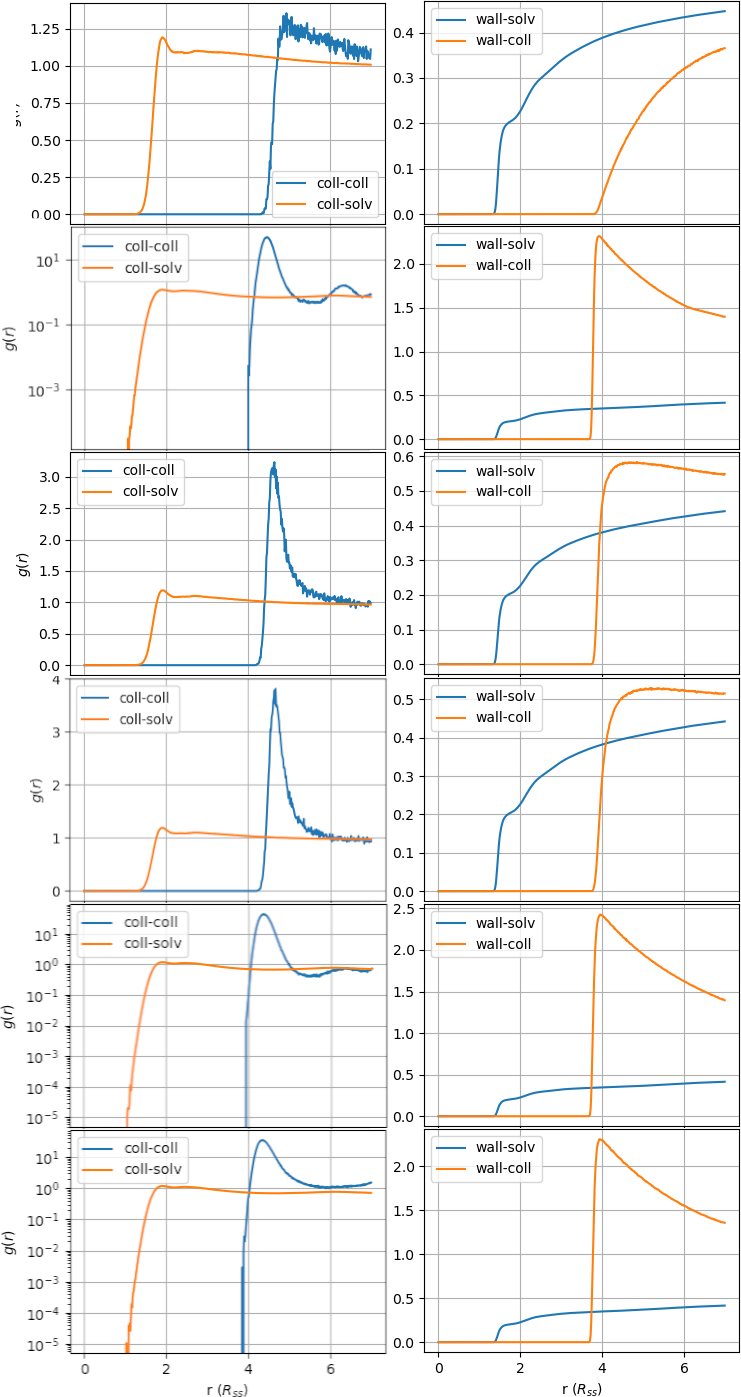}
    \caption{Radial distribution functions. \textbf{Left}: for colloids. \textbf{Right}: for walls. \\
    \textbf{1st row:} passive case without gravity. \textbf{2nd row:} passive case for $F_g=50$.\\
    \textbf{3rd row:} Pullers $F_p=16.67$ case for $F_g=0.0$. \textbf{4th row:} Pushers $F_p=16.67$ case for $F_g=0.0$.\\
    \textbf{5th row:} Pullers $F_p=33.33$ case for $F_g=50$. \textbf{6th row:} Pushers $F_p=33.33$ case for $F_g=50$.}
    \label{fig:enter-label}
\end{figure}

\clearpage

\bibliography{GW}